\documentclass[journal]{IEEEtran}
\usepackage{graphicx}
\usepackage{amsmath}
\usepackage{bm}
\usepackage{footnote}
\usepackage{amssymb}
\usepackage{subfigure} 
\usepackage{hyperref}
\usepackage{cite}
\usepackage{algorithm}
\usepackage{algorithm,algcompatible,amsmath}
\usepackage{algpseudocode}
\usepackage{xcolor}
\usepackage{cuted}
\usepackage{extarrows}
\usepackage{stmaryrd}

\newtheorem{theorem}{Theorem}[section]

\newtheorem{lemma}[theorem]{Lemma}
\newtheorem{proposition}[theorem]{Proposition}

\ifCLASSINFOpdf
\else
\fi

\newcommand{\mtight}{\setlength{\thickmuskip}{2mu} \setlength{\medmuskip}{2mu} \setlength{\thinmuskip}{2mu}}

\begin{document}

\title{Gridless DOA Estimation with Multiple Frequencies}

\author{Yifan Wu,
        Michael B. Wakin, \IEEEmembership{Fellow, IEEE}, and
        Peter Gerstoft, \IEEEmembership{Fellow, IEEE}
        
\thanks{Yifan Wu and Peter Gerstoft are with University of California, San Diego, La Jolla, CA, 92093, USA (e-mails:\{yiw062, pgerstoft\}@ucsd.edu).
}
\thanks{Michael. B. Wakin is with Colorado School of Mines, Golden, CO, 80401, USA (e-mail: mwakin@mines.edu)}
\thanks{Manuscript received \today}}

\maketitle

\begin{abstract}
Direction-of-arrival (DOA) estimation is widely applied in acoustic source localization. A multi-frequency model is suitable for characterizing the broadband structure in acoustic signals. In this paper, the \textit{continuous} (gridless) DOA estimation problem with multiple frequencies is considered. This problem is formulated as an atomic norm minimization (ANM) problem. The ANM problem is equivalent to a semi-definite program (SDP) which can be solved by an off-the-shelf SDP solver. The dual certificate condition is provided to certify the optimality of the SDP solution so that the sources can be localized by finding the roots of a polynomial. We also construct the dual polynomial to satisfy the dual certificate condition and show that such a construction exists when the source amplitude has a uniform magnitude. In multi-frequency ANM, spatial aliasing of DOAs at higher frequencies can cause challenges. We discuss this issue extensively and propose a robust solution to combat aliasing.  
Numerical results support our theoretical findings and demonstrate the effectiveness of the proposed method.
\end{abstract}


\IEEEpeerreviewmaketitle

\section{Introduction}

\IEEEPARstart{L}{ine} spectrum estimation  is a fundamental problem in signal processing, and has many applications in direction-of-arrival (DOA) estimation in sensor array processing \cite{van2004optimum}, wideband channel estimation \cite{venugopal2017channel}, and modern imaging modalities \cite{wei2021sar}. In line spectrum estimation, the observed signal $x[n]$ is a superposition of $K$ complex sinusoids (i.e. $x[n] = \sum_{k = 1}^K c_ke^{-j2 \pi f_k n}$) and the goal is estimating the frequencies $f_k$ of these $K$ sinusoids. An important application of line spectrum estimation is DOA estimation \cite{van2004optimum}. For DOA estimation, we have $K$ plane waves from angles $\{ \theta_1, \dots, \theta_K \}$  impinging on an array with $N_m$ sensors. Due to differen propagation delays to each sensor, the received data is a sum of $K$ spatial sinusoid vectors  $[1\; \dots\; e^{-j \frac{2 \pi f_0 (N_m - 1) d \cos \theta_k}{c}} ]^T (k \in \{ 1, \dots, K\})$ parameterized by the plane wave directions $\theta_k$ (\textcolor{black}{$f_0$} is a temporal frequency). Our goal is to estimate the $K$ DOAs ($\theta_k$) based on the received data. The cosine of each  DOA linearly maps to a single spatial frequency $\frac{2 \pi f_0 d \cos \theta_k}{c}$ of the sinusoid, and once the spatial frequencies are estimated, the DOA can be  retrieved. Many line spectrum estimation methods as multiple signal classification (MUSIC) \cite{schmidt1986multiple}, and estimation of signal parameters via rotational invariant techniques (ESPRIT) \cite{roy1989esprit}, have been used for narrow band signals.

Unfortunately, the aforementioned methods cannot be applied in wideband DOA estimation problems such as ocean acoustics localization and speaker localization. Wideband signal DOA estimation has been studied for decades \cite{wax1984spatio, wang1985coherent, buckley1988broad, di2001waves, yoon2006tops}. A subspace-based wideband DOA estimation approach, incoherent signal subspace method \cite{wax1984spatio}, was proposed with later improvement in the coherent signal subspace method (CSSM) \cite{wang1985coherent}. A broadband spatial-spectrum estimation approach \cite{buckley1988broad}  overcame the peak bias and source spectral content sensitivity from CSSM. Variants of CSSM, such as the weighted average of signal subspaces method \cite{di2001waves}, and the test of orthogonality of projected subspaces method \cite{yoon2006tops} were also proposed. Recently, some wideband DOA estimation methods based on sparse recovery have also been developed \cite{zhang2013wideband, wang2015novel, liu2011broadband, tang2011aliasing}. These sparsity-based methods have demonstrated superior performance compared to conventional methods. 

The multi-frequency (or multi-dictionary) model~\cite{tang2011aliasing,gemba2017, antonello2019joint,nannuru2019sparse, gemba2019} has shown success in modeling wideband signals. The multi-frequency model uses $N_f$ (rather than $1$) temporal frequency bins in a frequency set $\mathcal{F} = \{f_1, \dots, f_{N_f} \}$ to characterize a wideband signal. These frequencies are then used for estimation, as opposed to using a single  frequency under the narrowband model. The multi-frequency model was used for ocean acoustics localization \cite{gemba2019}. Most of the existing methods assume that the true spatial frequencies lie on a finite set of grid points, and their performance may degrade if the true spatial frequencies fall off the grid. 

To overcome the grid mismatch problem, atomic norm minimization (ANM) methods that work on continuous (gridless) dictionaries have been proposed in a variety of contexts \cite{chandrasekaran2012convex, candes2014towards, tang2013compressed, chi2014compressive, li2015off, fernandez2016super,chi2016guaranteed,yang2016exact, yang2016super, li2018atomic,  yang2015enhancing, yang2018sample, wagner2021, park2022gridless}. ANM extends grid-based, sparsity-promoting $\ell_1$ norm minimization to the continuous setting and is commonly applied to solve the line spectrum estimation problem for signals that are sparse in the temporal frequency domain. ANM was initially proposed in \cite{chandrasekaran2012convex}, which provides a general recipe for finding convex solutions to promote sparse decompositions, where one seeks to represent a given signal based on a minimal number of atoms from an atomic set composed of an ensemble of signal atoms. The ANM framework overcomes the grid mismatch issue and can achieve potentially infinite precision. However, all prior ANM works used a narrowband assumption and are not applicable for wideband DOA estimation.

\subsection{Related Work}
\subsubsection{Multiple Frequencies}
Multiple frequencies decompose a wideband signal into multiple narrowband signals and therefore are widely applied in acoustics source localization \cite{antonello2019joint, gemba2017, gemba2019} when the signal contains a wide range of frequency bins and cannot be characterized by a narrowband model. Some grid-based sparse localization approaches for the multiple frequencies were proposed \cite{gemba2017, gemba2019, tang2011aliasing, Liu2012,  nannuru2019sparse} for robustness and aliasing suppression.

\subsubsection{Atomic Norm Minimization}
ANM was initially proposed in \cite{chandrasekaran2012convex} as a general framework for promoting sparse signal decompositions. The pioneering ANM paper \cite{candes2014towards} worked directly with the continuous  (temporal) frequency estimation problem and considered the complete data case. As long as the temporal frequency separation was greater than a certain minimum separation, exact recovery of the active temporal frequencies was guaranteed. Furthermore, a semidefinite programming (SDP) framework that characterized the ANM problem was presented. The authors in \cite{tang2013compressed} studied continuous temporal frequency estimation based on randomly sampled data for the single measurement vector (SMV) case. The minimum separation condition was relaxed in \cite{fernandez2016super}. ANM for multiple measurement vectors (MMVs) was studied in \cite{li2015off, yang2016exact,  yang2018sample}. In \cite{chi2016guaranteed}, the author considered a super-resolution problem that had a similar setup to \cite{candes2014towards} except that the point spread function was assumed to be unknown. Based on the assumption that the point spread function was stationary and lived in a known subspace, the lifting trick was applied, and the problem was formulated using ANM. The model was generalized to non-stationary point spread functions in \cite{yang2016super}. The sample complexity of modal analysis with random temporal compression was established in \cite{li2018atomic}. ANM for 2D temporal frequency estimation was studied in \cite{chi2014compressive}. In \cite{yang2015enhancing}, the authors proposed a reweighted ANM framework, which enhances the sparsity and achieves super-resolution. An atomic norm for DOA estimation under gain-phase noise \cite{chen2020new} was proposed to mitigate the artifacts for electromagnetic signals. ANM was also recently applied in digital beamforming \cite{li2021digital, xenaki2015}, adaptive interference cancellation \cite{li2020adaptive}, denoising \cite{bhaskar2013atomic, li2019atomic}, and blind demodulation \cite{xie2019simultaneous, xie2020support}. We refer readers to \cite{chi2020harnessing} for a comprehensive overview of ANM and its applications. 

Our multi-frequency problem is different from the MMV problems~\cite{li2015off, yang2016exact,  yang2018sample} extensively studied in the past few years. Although both our work and MMVs fall under the general topic of multi-channel line spectrum estimation, the temporal frequencies in each channel are different in our problem while they are the same in MMVs. Therefore, each channel is modulated with a different sinusoid while this \textit{heterogeneous} modulation is absent in MMVs. This heterogeneous modulation leads to several challenges for theoretical analysis. First, it makes it difficult to derive an equivalent SDP problem based on the Vandermonde decomposition as has been done in many prior ANM works. Second, under our setup, each frequency other than the first will experience spatial aliasing of the DOAs. This leads to potential collisions or near collisions of the DOAs which are challenging to resolve. Thus, although having multiple frequencies does provide more data, one must ensure that aliasing does not undermine this benefit. These challenges make our problem more difficult to analyse than MMV problems. We will elaborate on these two challenges and our solutions in Sec. \ref{contribution}.

\subsection{Our Contributions}
\label{contribution}
In this work, we extend ANM to the multi-frequency framework so that it can be used for DOA estimation with wideband signals. Our contributions are summarized as follows:

\textbf{(1) Formulate an equivalent SDP problem.} Although ANM itself is a convex optimization problem, it is not directly solvable due to an infinite number of optimization parameters. Therefore, it is critical to find a computationally feasible solution that equivalently characterizes the ANM problem. Several prior works showed that certain ANM problems could be equivalently characterized by SDPs~\cite{tang2013compressed, li2015off, yang2016exact}. The derivation of an SDP problem typically relies on a Vandermonde decomposition, and equivalence with the ANM can be proved by showing that the SDP solution is both an upper and a lower bound for the ANM~\cite{tang2013compressed, li2015off, yang2016exact}. Unfortunately,  this commonly used technique cannot be applied in our case due to the heterogeneous temporal frequencies across different channels. In \cite{li2021digital, helland2019atomic}, certain SDPs were derived using the Vandermonde decomposition, but only the lower bound for the ANM problem could be guaranteed. In this work, we derive an equivalent SDP  based on the bounded real lemma for trigonometric polynomials~\cite{dumitrescu2017positive}. This equivalent SDP will provide a computationally feasible solution for the ANM when multiple frequencies are considered. We also explain how our SDP is the dual to a minor adaptation of the SDP proposed in~\cite{yang2017gridless} for line spectrum estimation with harmonics.

\textbf{(2) Provide the dual certificate condition.} We derive a dual certificate condition that can be used to certify the optimal atomic decomposition. In particular, the DOAs of the sources are localized with the help of the dual polynomial arising from the ANM optimization problem. As long as the dual polynomial satisfies the dual certificate condition, the frequencies can be localized by finding the roots of a polynomial. Therefore, the dual certificate condition not only provides a theoretical guarantee for the optimality, but also offers a method for the DOA estimation. 

\textbf{(3) Construct the dual polynomial that satisfies the dual certificate condition.} In cases where we can prove the existence of a dual polynomial that satisfies the dual certificate condition, then the optimality and therefore exact DOA estimation are guaranteed. If the array spacing $d \leq \frac{c}{2N_ff_0} = \frac{\lambda_{N_f}}{2}$, spatial aliasing would be fully avoided for all of the temporal frequencies, and it may be possible to construct a valid dual polynomial under a mild separation assumption on the source directions. In such a case, the success of the algorithm is guaranteed. 

The dual polynomial is developed our model for arbitrary spacing $d$. A larger aperture $(N_m-1)d$ with greater $d$ may improve spatial resolution but introduces spatial aliasing. If the spacing $d = \frac{c}{2f_0} = \frac{\lambda_1}{2}$, spatial aliasing is present in all but the first frequency. This spacing necessarily creates periodicity in all but the first frequency of the vector-valued dual polynomial. 
Such periodicity brings the risk of creating ambiguity in the source direction.
More specifically, after spatial aliasing, when two source directions coincide at one frequency, we refer to this as  \textit{collision}. 
Collision may happen in multiple frequency bins, and  it becomes more likely for great $N_f$. 
Most ANM works need well-separated harmonics to work~\cite{candes2014towards, tang2013compressed, li2015off, yang2016exact}. However, in a multi-frequency scenario, one must consider the separations for DOAs across all frequencies.
Assuming collisions and near collisions are thus avoided and under some additional assumptions about the source amplitudes, we guarantee that there exists a dual polynomial satisfying the dual certificate condition.

\textbf{(4) Implementation.} We propose a fast implementation so that the SDP has a reduced size. This fast implementation also extends the approach to an arbitrary set of frequencies. Numerical results show that the dual polynomial still serves as a precise indicator for the DOAs. Hence, in terms of the DOA estimation, the algorithm succeeds even when collisions are present. 


Finally, our work is inspired by recent advances in ANM for super-resolution, but significantly deviates from the existing MMV works. This work significantly extends our previous ICASSP paper \cite{wu2022gridless}. It includes additional analysis for the dual polynomial construction, aliasing and collision, and provides a fast algorithm and extensive simulations. This paper is the first work that extends ANM to multiple frequencies so that it can be adapted to \textit{gridless} DOA estimation for wideband signals via convex programming. 

\subsection{Notations and Organization}
 Boldface letters  represents matrices and vectors. Conventional notations $(\cdot)^T$, $(\cdot)^H$, $(\cdot)^*$, $\langle \cdot \rangle_{\mathbb{R}}$, and $\langle \cdot \rangle$ stand for matrix/vector transpose, Hermitian transpose, complex conjugate, real inner product, and inner product, respectively. $\mathrm{Tr}(\cdot)$ is used to represent the trace of a matrix. $\| \cdot \|_p$ and $\| \cdot \|_F$ are used to express vector $\ell_p$ norm and matrix Frobenius norm. For a Hermitian matrix $\mathbf{A}$, $\mathbf{A} \succeq 0 $ means $\mathbf{A}$ is a positive semidefinite (PSD) matrix. $\odot$ stands for the Hadamard product.  The $\ell_{1, 2}$ norm of a matrix $\mathbf{A} = [\mathbf{a}_1 \; \dots \; \mathbf{a}_N]$ is defined as $\|\mathbf{A}\|_{1, 2}:= \sum_{i = 1}^{N} \|\mathbf{a}_i\|_2$. The imaginary unit is denoted by $j = \sqrt{-1}$.

The rest of the paper is organized as follows. Sec. \ref{sec:2} introduces the signal model and the assumptions. The equivalent SDP and the dual certificate condition are derived in Sec. \ref{sec:3}. Sec. \ref{sec:4} constructs the dual polynomial that satisfies the dual certificate condition and also analyses the collision and near collision issues. Sec. \ref{sec:5} presents some numerical examples to support and demonstrate theoretical findings. Finally, Sec. \ref{sec:6} concludes the paper. 

\section{\label{sec:2} Signal Model}
\label{model}
\subsection{Assumptions and Model Framework}
\subsubsection{Assumptions}
The following assumptions are made for the array configuration and signal model:
\begin{enumerate}
    \item There are $N_m$ sensors forming a uniform linear array (ULA) with array spacing $d$.
    \item There are $K$ active sources impinging on the array from unknown directions of arrival (DOAs) $\theta$.
    \item Each source has $N_f$ active temporal frequency components, each at an integer multiple $f$ of a fundamental frequency $f_0$, i.e., $f \in \{1, \dots, N_f\}$ and $ff_0 \in \{f_0, \dots, N_ff_0\}$. This is only a technical assumption to simplify the analysis, and our method can be applied in any frequency set with the fast algorithm proposed in Sec. \ref{fast_ag}.
    \item Suppose $d \leq \frac{c}{2f_0}$ holds (or, equivalently, $\frac{2\pi f_0d}{c} \leq \pi$),
    where $c$ is the speed of propagation. We also notice that $d = \frac{c}{2f_0}$ is the maximum separation to avoid spatial aliasing at the fundamental frequency. For higher frequencies (i.e. $f \geq 2$), aliasing will still exist. Such aliasing is not considered in conventional narrowband ANM papers. It is possible to develop the method with $d = \frac{c}{2N_ff_0}$ so that aliasing can be completely avoided in all frequencies. 
    
\end{enumerate}

\subsubsection{Multiple Frequencies}
Based on the above assumptions, we absorb the constant parameters $d$, $f_0$, and $c$ into a scaled DOA parameter $w = w(\theta) := \frac{f_0d \cos(\theta)}{c} \in [-f_0d/c, f_0d/c]$. Henceforth, $w$ is simply referred as the DOA. 

\begin{figure}[t]
\centering
\includegraphics[width=8.5cm, height = 5cm]{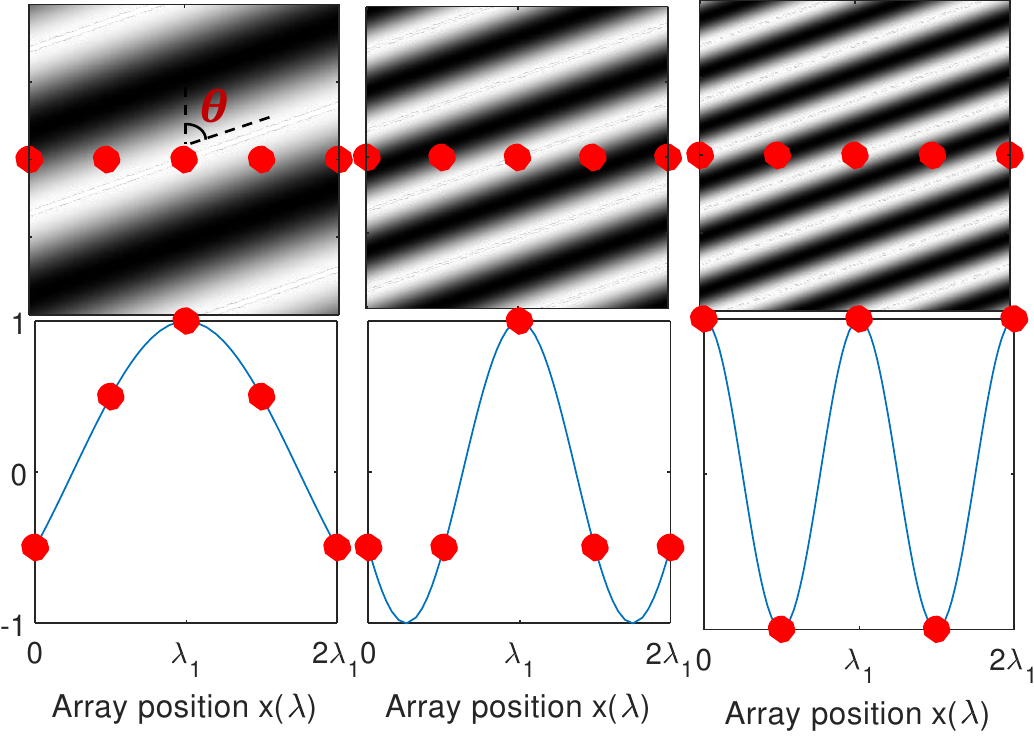}
\caption{Multi-frequency data on array with $N_m = 5$ sensors. Top row: time snapshot of propagating plane wave with angle of arrival $\theta$ and temporal frequency (left to right) $f_0$, $2f_0$, $3f_0$. Bottom row: array data are samples of a spatial sinusoid whose spatial frequency depends on the temporal frequency and DOA. Only the real part of the array data is shown. 
}
\label{tf_sf}
\end{figure}

For each temporal frequency $ff_0 \in \{1, \dots, N_f\} \cdot f_0$, let $\mathbf{y}_f \in \mathbb{C}^{N_m}$ denote the received signal across the $N_m$ sensors. $\mathbf{y}_f$ can be expressed as a sum of $K$ spatial sinusoid vectors, with the $k$-th vector having spatial frequency $fw(\theta_k)$. Importantly, the spatial frequency depends on both the temporal frequency $ff_0$ and the DOA $w(\theta_k)$. To better illustrate these effects, we refer the reader to Fig. \ref{tf_sf}. Suppose $N_f = 3$, $N_m = 5$, and the input signal (top row) is a complex sinusoid with temporal frequency $ff_0$. The spatial samples obtained from the sensors (red) will be sampled sinusoids (bottom row) with different spatial frequencies that depend on both the temporal frequency and the DOA. 

Stacking all of the data from the $N_f$ frequencies into a matrix, the full set of received data is denoted by $\textbf{Y} := [\mathbf{y}_1\; \dots\; \mathbf{y}_{N_f}]  \in \mathbb{C}^{N_m \times N_f}$. Summing over the $K$ active DOAs, we write
\begin{equation}
\label{y_signal}
\begin{aligned}
     &\mathbf{Y} = \mathbf{X} + \mathbf{W}, 
\end{aligned}
\end{equation}
where
\begin{equation}
\label{KR}
\begin{aligned}
\mathbf{X} &:= \sum_{w} c_w [x_w(1)\mathbf{a}(1, w) \; \dots \; x_w(N_f)\mathbf{a}(N_f, w)]  \\
    &= \sum_{w} c_w \mathbf{A}(w) \varoast \mathbf{x}_w^{T},
\end{aligned}
\end{equation}
$\mathbf{a}(f, w)  := [1 \;\dots \; e^{-j2\pi w f(N_m - 1)} ]^T  \in \mathbb{C}^{N_m}$ is the array manifold vector (steering vector) corresponding to the $f$-th frequency bin and DOA $w$, $x_w(f)$ is the signal amplitude for the $f$-th frequency bin, and $\mathbf{W} := [\mathbf{w}_1, \dots, \mathbf{w}_{N_f}] \in \mathbb{C}^{N_m \times N_f}$ is additive Gaussian uncorrelated noise. $\mathbf{x}_w := [x_w(1) \; \dots \; x_w(N_f)]^T \in \mathbb{C}^{N_f}$ is a collection of $N_f$ amplitudes corresponding to the same DOA, $\mathbf{A}(w) := [\mathbf{a}(1, w) \; \dots \; \mathbf{a}(N_f, w)] \in \mathbb{C}^{N_m \times N_f}$, and $\varoast$ is the Khatri-Rao product defined as $[\mathbf{A}(w) \varoast \mathbf{x}_w^T] := [\mathbf{a}(1, w)x_w(1) \; \dots \; \mathbf{a}(N_f, w)x_w(N_f)] \in \mathbb{C}^{N_m \times N_f}$. We assume that $\|\mathbf{x}_w\|_2 = 1$; the coefficient $c_w$ absorbs any other scaling of the source amplitudes $c_w\mathbf{x}_w$. Our goal is to identify the $K$ active DOAs $w$ from the data matrix $\mathbf{Y}$. 

In the following \textcolor{black}{sections}, we  primarily develop the optimization methods within the noise-free model, i.e., where $\mathbf{W = 0}$. We describe, however, how the optimization problem is modified if noise is present (see~\eqref{ANM_noise} and \eqref{ssdp_robust}).

\subsection{Mapping Operator}
In this section, we will define some mapping operators that help us set up our method.  Define
\begin{equation}
\begin{aligned}
     \mathbf{z} = \mathbf{z}(w) &:= [1 \quad e^{-j2 \pi w1} \dots e^{-j2 \pi wN_f(N_m-1)}]^T  \\
     &= [z^0 \quad z^1 \dots z^{N-1}]^T \in \mathbb{C}^{N}
\end{aligned}
\end{equation}
that collects all possible complex exponentials from the array manifolds in all frequencies, where $N := N_f(N_m - 1)+1$ and $z = z(w) := e^{-j2 \pi w}$. The intuition for defining the $\mathbf{z}$ notation will be explained after the definition of the dual polynomial vector~\eqref{dual_poly_def}. Introduce $\mathbf{Z}=\mathbf{Z}(w) := [\mathbf{z} \dots \mathbf{z}] \in \mathbb{C}^{N \times N_f}$ and define $\mathbf{X}' \in \mathbb{C}^{N \times N_f}$ as
\begin{equation}
\label{X'}
    \mathbf{X'} := \bigg(\sum_{w} c_w \mathbf{Z} \varoast \mathbf{x}_w^{T}\bigg).
\end{equation}

\begin{figure}[t]
\centering
\includegraphics[width=6.5cm, height = 4cm]{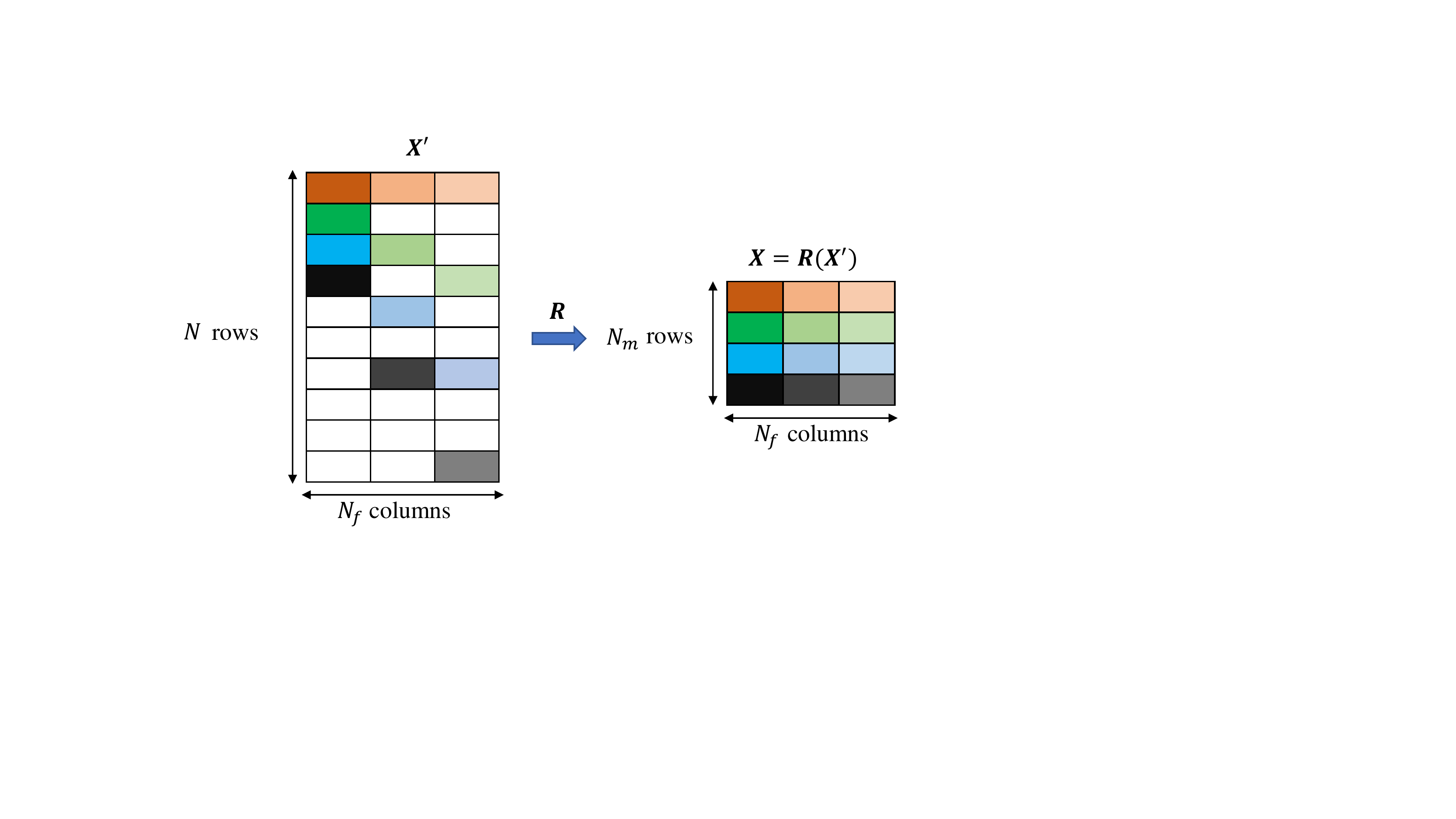}
\caption{Compaction of matrix $\mathbf{X}'$ to $\mathbf{X}$ by mapping $\mathcal{R}: N \times N_f \rightarrow N_m \times N_f$ defined in \eqref{R_mapping}. }
\label{R_map_fig}
\end{figure}

Then, we define the $\mathcal{R}$ operator that maps $\mathbf{X}'$ to $\mathbf{X}$ as 
\begin{equation}
\label{R_mapping}
 \mathbf{X} = \mathcal{R}(\mathbf{X'}) \Rightarrow \mathbf{X}(i, j) = \mathbf{X}'(1+(i-1)j, j),
\end{equation}
where $\mathcal{R}: N \times N_f \rightarrow N_m \times N_f$ is a mapping that selects $N_m$ elements from the $N$ elements in each column of $\mathbf{X}'$. We demonstrate the mapping in Fig. \ref{R_map_fig}. 
Note that $\mathbf{A}(w)$ can be represented in terms of $\mathbf{Z}$ by using the $\mathcal{R}$ operator as
\begin{equation}
\label{Z_A}
\mathbf{A}(w) = \mathcal{R}(\mathbf{Z}).
\end{equation}

Note that in \cite[(8)]{yang2017gridless}, an analogous mapping operator is introduced in the context of the line spectrum estimation problem with harmonics. In \cite[(9)]{yang2017gridless}, the transformation is applied in the signal space and enables the formulation of an SDP problem in the primal domain. In our paper, $\mathcal{R}$ is applied to the coefficient matrix $\mathbf{H}$ (see \eqref{R_dual}) and that enables us to formulate an SDP problem in the dual domain.

\section{\label{sec:3}Methodology}

\subsection{Atomic Norm Minimization (ANM)}
To efficiently represent matrices of the form~\eqref{KR}, we define the atomic set 
\begin{equation}
\label{atom}
    \mathcal{A} := \{\mathbf{A}(w) \varoast \mathbf{x}_w^T: \;  w \in [-f_0d/c, f_0d/c],  \|\mathbf{x}_w\|_2 = 1 \}.
\end{equation}
From (\ref{KR}), $\mathbf{X}$ is a sparse combination of $K$ atoms from $\mathcal{A}$ since only a few directions have active sources. ANM provides a framework for identifying such sparse combinations in continuously parameterized dictionaries. In our case, the dictionary $\mathcal{A}$ is parameterized by the continuous DOA $w$. 

In the the noise-free case, to identify the $K$ active DOAs $w$ from the data matrix $\mathbf{Y}$, we propose the following ANM-based optimization framework:
\begin{equation}
\label{ANM}
\begin{aligned}
    \min_{\mathbf{X}} \quad & \|\mathbf{X}\|_{\mathcal{A}} \quad  \textrm{s.t.}   \quad & \mathbf{Y} = \mathbf{X},
\end{aligned}
\end{equation}
where the atomic norm  is defined as 
\begin{equation}
\label{anm_def}
\begin{aligned}
\|\mathbf{X}\|_{\mathcal{A}} &:= \inf \{t \geq 0| \mathbf{X} \in t \cdot \text{conv}(\mathcal{A}) \} \\
&= \inf \big\{\sum_w |c_w| \big|  \mathbf{X} = \sum_w c_w\mathbf{A}(w) \varoast \mathbf{x}_w^T\big\}.
\end{aligned}
\end{equation}

When noise is present, we modify the optimization problem to relax the equality constraint:
\begin{equation}
\label{ANM_noise}
\begin{aligned}
    \min_{\mathbf{X}} \quad & \|\mathbf{X}\|_{\mathcal{A}} \quad  \textrm{s.t.}   \quad & \|\mathbf{Y} - \mathbf{X}\|_F \leq \eta,
\end{aligned}
\end{equation}
where $\eta$ depends on the noise level.

It is not obvious how to obtain DOAs directly based on \eqref{ANM} (and \eqref{ANM_noise}), as one of the solutions is $\mathbf{Y}$ itself. In the following sections, we develop an equivalent optimization problem for computing the atomic decomposition of $\mathbf{Y}$, which enables determining the DOAs via the dual polynomial.

\subsection{Dual Atomic Norm and Dual Polynomial}

Let $\|\mathbf{X}\|$ be a matrix norm. The associate \textit{dual norm}, denoted $\|\mathbf{Q} \|^*$, is defined as \cite[Appendix A.1.6]{boyd2004convex}, 
\begin{equation}
\label{dual_norm}
    \|\mathbf{Q}\|^* := \sup_{\|\mathbf{X}\|\leq 1} \langle \mathbf{Q}, \mathbf{X} \rangle_{\mathbb{R}}.
\end{equation}
Also note that the dual of the dual norm is the primal norm.

Now we apply \eqref{dual_norm} to the atomic norm.   The primal atomic norm $\|\mathbf{X}\|_{\mathcal{A}}$ is expressed in terms of the dual atomic norm $\|\mathbf{Q}\|^{*}_{\mathcal{A}}$ (where $\mathbf{Q} := [\mathbf{q}_1 \dots \mathbf{q}_{N_f}] \in \mathbb{C}^{N_m \times N_f}$ is the dual variable) as 
\begin{equation}
\label{norm}
\begin{aligned}
    \|\mathbf{X}\|_{\mathcal{A}} &:= \sup_{\|\mathbf{Q}\|^{*}_{\mathcal{A} }\leq 1} \langle \mathbf{Q}, \mathbf{X} \rangle_{\mathbb{R}} = \sup_{\|\mathbf{Q}\|^{*}_{\mathcal{A} }\leq 1} \langle \mathbf{Q}, \mathbf{Y} \rangle_{\mathbb{R}}, 
\end{aligned}
\end{equation}
where the last equality is only for the noise-free case (see the constraint in \eqref{ANM}).

For any dual variable $\mathbf{Q}$, we define the corresponding {\em dual polynomial vector} $\bm{\psi}(\mathbf{Q}, w) \in \mathbb{C}^{N_f}$ as 
\begin{equation}
\label{dual_poly_def}
\begin{aligned}
     \bm{\psi}(\mathbf{Q}, w)&:= [\mathbf{q}_{1}^{H}\mathbf{a}(1, w) \dots \mathbf{q}_{N_f}^{H}\mathbf{a}(N_f, w)]^T \\
     &= [\sum_{m = 1}^{N_m}q_1^*(m)z^{(m-1)} \dots \sum_{m = 1}^{N_m}q_{N_f}^*(m)z^{N_f \cdot (m-1)} ]^T.
\end{aligned}
\end{equation}
Note that each entry in $\bm{\psi}(\mathbf{Q}, w)$ is a polynomial in $z$. The dual polynomial will be useful for setting up the dual certificate condition and extracting the DOA (see Sec. \ref{dual_cert} and Sec. \ref{DOA_extract}). However, since each frequency has different array manifold vectors, it is difficult to express $\bm{\psi}(\mathbf{Q}, w)$ as a matrix product of $\mathbf{Q}$ and a vector. To construct a homogeneous representation for $\bm{\psi}(\mathbf{Q}, w)$, we will leverage $\mathbf{z}$, an ensemble of the array manifold, and the matrix $\mathbf{H} \in \mathbb{C}^{N \times N_f}$ defined in terms of $\mathbf{Q}$ as follows ($m = \{1, \dots, N_m \}, f = \{1, \dots, N_f \}$)
\begin{equation}
\label{Hij_new}
    \mathbf{H}(i, f) = \left\{
\begin{array}{ll}
    \mathbf{Q}(m, f) \quad   &\mbox{for} \;\; (i,f)=(f \cdot (m-1) + 1, f)\\
    0  & \textrm{otherwise}, 
\end{array}
\right.
\end{equation}
or $\mathbf{H} = \mathcal{R}^*(\mathbf{Q})$, where $\mathcal{R}^*:N_m \times N_f \rightarrow N \times N_f$ is the adjoint mapping of $\mathcal{R}$.
Note the relationship between $\mathbf{Q}$ and $\mathbf{H}$ can be alternatively expressed as 
\begin{equation}
\label{R_dual}
    \mathbf{Q} = \mathcal{R}(\mathbf{H}).
\end{equation}

With the help of $\mathbf{H}$ and $\mathbf{z}$, $\bm{\psi}(\mathbf{Q}, w)$ has the homogeneous representation 
\begin{equation}
\label{dual_homo}
\bm{\psi}(\mathbf{Q}, w) = \mathbf{H}^H\mathbf{z}.
\end{equation}

Now, we consider  $\|\mathbf{Q}\|^{*}_{\mathcal{A}}$, which appears in a constraint in \eqref{norm}. Recalling that $\|\mathbf{x}_w\|_2$ = 1, we have
\begin{equation}
\label{anmdual}
\begin{aligned}
\|\mathbf{Q}\|^{*}_{\mathcal{A}} &:= \sup_{\|\mathbf{X}\|_{\mathcal{A}} \leq 1} \langle \mathbf{Q, X} \rangle_{\mathbb{R}} = \sup_{\|\mathbf{X}\|_{\mathcal{A}} \leq 1}\text{Re}[\mathrm{Tr}(\mathbf{Q}^H\mathbf{X})]\\
&= \sup_{\substack{\mathbf{x}_w \\ w}} \text{Re}[\mathrm{Tr}(\mathbf{Q}^H\mathbf{A}(w) \varoast \mathbf{x}_w^T)]      \\
&= \sup_{\substack{x_w(f) \\ w}} \text{Re}\bigg(\sum_{f=1}^{N_f} x_w(f)\mathbf{q}_{f}^{H}\mathbf{a}(f, w)\bigg) \\
&\overset{(\text{a})}= \sup_{\substack{\mathbf{x}_w \\ w}} \text{Re} (\mathbf{x}_w^{H}\bm{\psi}(\mathbf{Q}, w)) = \sup_{\substack{\mathbf{x}_w \\ w}} |\mathbf{x}_w^{H}\bm{\psi}(\mathbf{Q}, w)|\\
&\overset{(\text{b})}= \sup_{w} \|\bm{\psi}(\mathbf{Q}, w)\|_2  = \sup_{w} \|\mathbf{H}^H\mathbf{z}\|_2
\end{aligned}
\end{equation}
where (a) follows by the definition of the dual polynomial vector and (b) follows from the definition of the operator norm.

Using \eqref{anmdual}, the condition  $\|\mathbf{Q}\|^{*}_{\mathcal{A}} \leq 1$ can be equivalently formulated as an SDP constraint. To simplify the theoretical analysis, we assume $d = \frac{c}{2f_0}$ and thus $w \in [-1/2, 1/2]$ here. We however notice that the ``if'' part can be generalized to any $d \leq \frac{c}{2f_0}$.

\begin{proposition}
\label{p1}
Let $\bm{\psi}(\mathbf{Q}, w)$ be as defined in (\ref{dual_poly_def}) and $w \in [-1/2, 1/2]$. Then $\|\mathbf{Q}\|^{*}_{\mathcal{A}}  \leq 1$ holds if and only if there exists a matrix $\mathbf{P}_0 \in \mathbb{C}^{N \times N} \succeq 0$ such that
\begin{equation}
\label{tr}
      \sum_{i = 1}^{N-k} \mathbf{P}_{0} (i, i + k) = \delta_k = \left\{
\begin{array}{cc}
\begin{aligned}
  & 1, \quad k = 0,\\
  & 0, \quad k = 1, \dots, N-1,
\end{aligned}
\end{array}
\right.
\end{equation}
and such that 
\begin{equation}       
\label{sdp}
\left[                 
  \begin{array}{cc}   
    \mathbf{P}_0 & \mathbf{H} \\
    \mathbf{H}^H & \mathbf{I}_{N_f} \\  
  \end{array}
\right]  \succeq 0.             
\end{equation}
\end{proposition}
\textit{Proof} See Appendix \ref{proof_3.1}. $\hfill\square$

\subsection{SDP Formulations of ANM Problems}
\subsubsection{Noise-free ANM}
In the noise-free case, based on Proposition~\ref{p1} and~\eqref{norm}, we have an SDP that is equivalent to (\ref{ANM}):

\begin{equation}
\label{ssdp}
\begin{aligned}
    & \max_{\mathbf{Q}, \mathbf{P}_0} \langle \mathbf{Q}, \mathbf{Y} \rangle_{\mathbb{R}} 
    \quad \textrm{s.t.}  \left[                 
  \begin{array}{cc}   
    \mathbf{P}_0 & \mathbf{H} \\  
    \mathbf{H}^H & \mathbf{I}_{N_f} \\  
  \end{array}
\right]  \succeq 0, \\
&\sum_{i = 1}^{N-k} \mathbf{P}_0(i, i+k) = \delta_k, \mathbf{H} = \mathcal{R}^*(\mathbf{Q}),
\end{aligned}
\end{equation}
where the dual variable $\mathbf{Q} \in \mathbb{C}^{N_m \times N_f}$, and $\mathbf{H}$ is related to $\mathbf{Q}$ as in~\eqref{Hij_new}.
\subsubsection{Robust ANM}
\label{sdp_collision}
To make ANM robust to noise and near collisions (see \eqref{near_collision}), we use the following alternative to \eqref{ssdp}:
\begin{align}
     &\max_{\mathbf{Q}, \mathbf{P}_0} \langle \mathbf{Q}, \mathbf{Y} \rangle_{\mathbb{R}}   - \eta\|\mathbf{Q}\|_{F} - \lambda\|\mathbf{Q}\|_{1, 2}
    \quad \textrm{s.t.} \left[                 
  \begin{array}{cc}   
    \mathbf{P}_0 & \mathbf{H} \\  
    \mathbf{H}^H & \mathbf{I}_{N_f} \\  
  \end{array}
\right]  \succeq 0, \nonumber\\
&\sum_{i = 1}^{N-k} \mathbf{P}_0(i, i+k) = \delta_k, \mathbf{H} = \mathcal{R}^*(\mathbf{Q}),\label{ssdp_robust}
\end{align}
where the term $\eta\|\mathbf{Q}\|_{F}$ suppresses noise  \cite[(15)]{chi2016guaranteed} \cite[(34), and App. D]{xenaki2015}. The value of $\eta$ is the same as in \eqref{ANM_noise} \cite{xenaki2015, chi2016guaranteed}. Based on similar arguments to \cite[App. D]{xenaki2015}, \eqref{ssdp_robust} with $\lambda = 0$ is the dual problem of \eqref{ANM_noise}. We further add an $\ell_{1, 2}$ regularization term to suppress near collisions. The $\ell_{1, 2}$ regularization term  $\lambda \|\mathbf{Q}\|_{1, 2}$ promotes column sparsity, and it reduces the contributions from the ``bad frequencies''. Near collision is a phenomenon that arises in our multi-frequency ANM model, and it is introduced in Sec. \ref{aliase_collision}. For the noise-free data, one may set $\eta = 0$, and for the near-collisions-free data, one may set $\lambda = 0$.

\subsection{Dual Certificate}
\label{dual_cert}
The dual polynomial $\bm{\psi}(\mathbf{Q}, w)$ introduced in \eqref{dual_poly_def} serves as a certificate for the optimality of (\ref{ANM}) and can therefore be used to extract the unknown DOAs. Specifically, we have the following dual certificate theorem, which is inspired by \cite[Proposition II.4]{tang2013compressed}. To ensure uniqueness, a linear independence assumption is added. 
\begin{theorem} Define $\mathcal{W} := \{w_1, \dots, w_K \}$ as a collection of DOAs with cardinality $K$. Then $\mathbf{X} = \sum_{w \in \mathcal{W}} c_w \mathbf{A}(w) \varoast \mathbf{x}_w^T$ \text{($\|\mathbf{x}_w\|_2 = 1$)} is the unique atomic decomposition such that $\|\mathbf{X}\|_{\mathcal{A}} = \sum_{w \in \mathcal{W}} |c_w|$ if the following two conditions are satisfied: \\
(1) There exists $\mathbf{Q}$ such that the dual polynomial vector $\bm{\psi}(\mathbf{Q}, w)$ satisfies 
\begin{equation}
\label{cert}
\left\{
\begin{array}{cc}
\begin{aligned}
   & \bm{\psi}(\mathbf{Q}, w) = \mathrm{sign}(c_w^*) \mathbf{x}_w  \quad \forall w \in \mathcal{W}\\
   & \|\bm{\psi}(\mathbf{Q}, w)\|_2 < 1 \quad \forall w \notin \mathcal{W},
\end{aligned}
\end{array}
\right.
\end{equation}
where $\mathrm{sign}(c_w^*) := \frac{c_w^*}{|c_w^*|}$. \\
(2) $\{ \mathbf{A}(w) \varoast \mathbf{x}_w^T: \; w \in \mathcal{W} \}$ is a linearly independent set.
\label{thm:cert}
\end{theorem}
\textit{Proof} See Appendix \ref{proof_3.2}. $\hfill\square$

\subsection{DOA Extraction}
\label{DOA_extract}
Based on Theorem 3.2, we know if \eqref{cert} is satisfied, the optimality is guaranteed. In (\ref{cert}), $\|\bm{\psi}(\mathbf{Q}, w)\|_2 = 1$ for $w \in \mathcal{W}$. After solving the SDPs \eqref{ssdp}--\eqref{ssdp_robust} by CVX \cite{grant2014cvx}, the optimal dual variables $\mathbf{Q}$ (and thus $\mathbf{H}$) are obtained. Then, the DOA is retrieved by finding the roots for $R(w)$ defined in \eqref{R(w)}.

\begin{figure}[t]
\centering
\includegraphics[width=8.5cm]{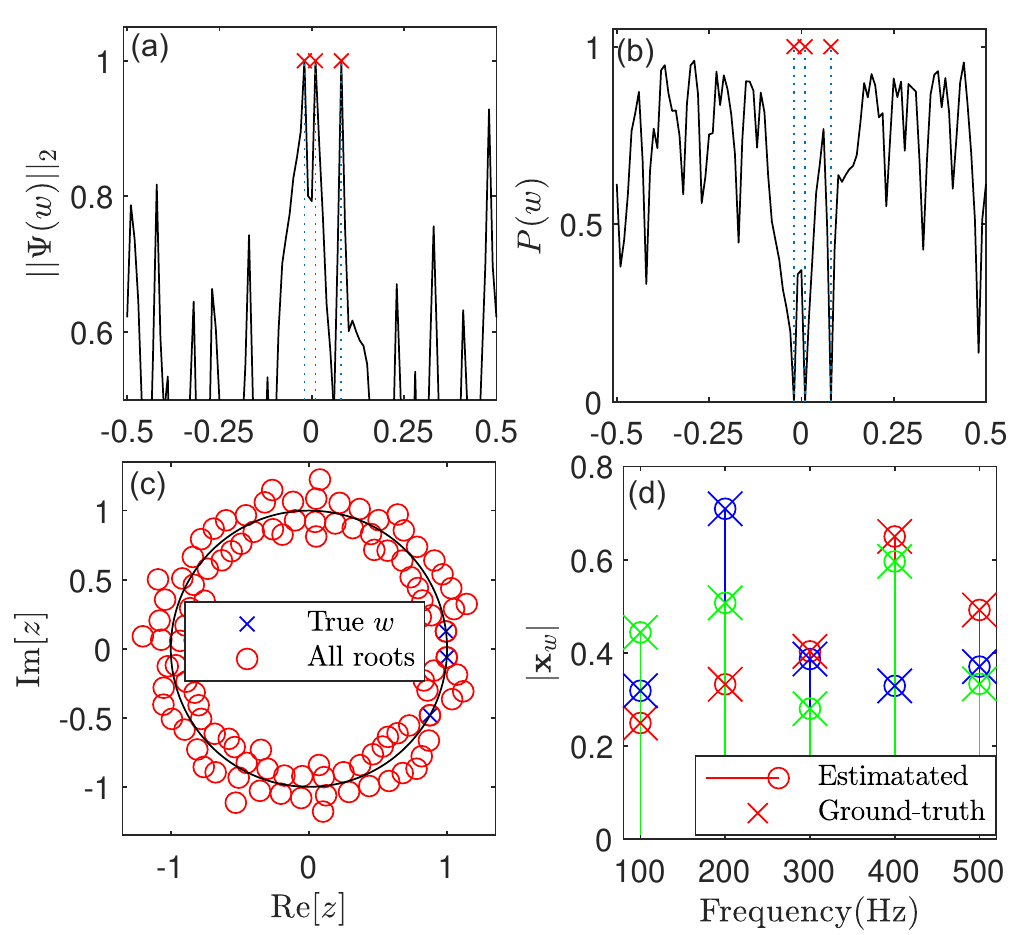}
\caption{DOA extraction through the dual polynomial. An ULA with $N_m = 12$ sensors, and spacing $d = c/2f_0$ is used. $N_f = 5$. $\theta = [80.7931^\circ, 88.854^\circ, 92.2924^\circ]$, and $w = [0.08, 0.01, -0.02]$. (a) $\|\bm{\psi}(\mathbf{Q}, w)\|_2$ versus $w$; (b) $P(w)$ versus $w$; (c) Roots for $P(w)$; (d) Amplitude estimation for each frequency (three colors are used to indicate different sources).}
\label{Root_demo}
\end{figure}

Based on \eqref{dual_homo}, $R(z)$ has the polynomial representation
\begin{equation}
\label{poly_trig}
    R(z) = 1 - \mathbf{z}^H \mathbf{P}_1 \mathbf{z} = 1 - \sum_{i = -(N - 1)}^{(N - 1)} r_i z^{i},
\end{equation}
where $\mathbf{P}_1 := \mathbf{H} \mathbf{H}^H$ and $r_k := \sum_{i = 1}^{N - k}\mathbf{P}_1(i, i + k)$. Indeed, $R(w)$ is a polynomial with degree $2(N - 1)$. The roots $\hat{z}$ can be obtained, and $\hat{w}$ is retrieved by locating the roots of $R(z)$ on the unit circle (see Fig. \ref{Root_demo} (c)):
\begin{equation}
    \hat{w} = \bigg\{-\frac{\angle \hat{z}}{2 \pi} \bigg| R(\hat{z}) = 0, |\hat{z}| = 1 \bigg\}.
\end{equation}
Note $\angle \hat{z} = -2\pi \hat{w} = -\frac{2 \pi f_0 d}{c} \cos \theta = \frac{2 \pi f_0 d}{c} \cos(\pi - \theta)$. $\hat{\theta}$ is therefore  estimated by
\begin{equation}
    \hat{\theta} = \pi - \cos^{-1}\bigg(\frac{\angle \hat{z}}{2 \pi f_0 d / c} \bigg).
\end{equation}

The implementation details for the proposed algorithm are summarized in Algorithm \ref{ag}.

\algnewcommand\INPUT{\item[\textbf{Input:}]}%
\algnewcommand\OUTPUT{\item[\textbf{Output:}]}%
\algnewcommand\Initialize{\item[\textbf{Initialization:}]}

\begin{algorithm} 
\caption{Gridless DOA estimation algorithm}\label{ag}
\begin{algorithmic}
\INPUT $\mathbf{Y} \in \mathbb{C}^{N_m \times N_f}, d, f_0, c, K$, $\eta$ (for noisy case), $\lambda$ (for near collision case) 
\Initialize 
\State (For noisy or near collision case) Solve \eqref{ssdp_robust} by CVX and obtain $\mathbf{H}$ 
\State (Otherwise) Solve \eqref{ssdp} by CVX and obtain $\mathbf{H}$ from $\mathbf{Q}$
\State $\mathbf{P}_1 \gets \mathbf{H}\mathbf{H}^H$
\State $N \gets N_f(N_m - 1) + 1$
\While{$-(N - 1) \leq k \leq (N - 1)$}

\State $r_k \gets \sum_{i = 1}^{N - k}\mathbf{P}_1(i, i + k)$

\EndWhile
\State $\mathbf{r} \gets [-r_{-(N - 1)} \dots -r_{(N - 1)}]$
\State $\mathbf{r}(N) \gets \mathbf{r}(N) + 1$
\State $\mathbf{roots} \gets \mathrm{roots}(\mathbf{r})$
\State [$\mathbf{dist, ind}$] $\gets$ $\mathrm{sort(abs(1 - abs(\mathbf{roots}))}$
\State $\mathbf{roots\_sort} \gets \mathbf{roots}(\mathbf{ind})$
\State $\mathbf{roots\_unique} \gets \mathbf{roots\_sort}(1:2:2K)$
\State $\hat{\mathbf{\theta}} \gets 180 - \mathrm{acosd(\mathrm{angle}(\mathbf{roots\_unique})} /(f_0 d / c))$

\OUTPUT $\hat{\mathbf{\theta}}$

\end{algorithmic}
\end{algorithm}

\begin{figure}[t]
\centering
\includegraphics[width=8.5cm]{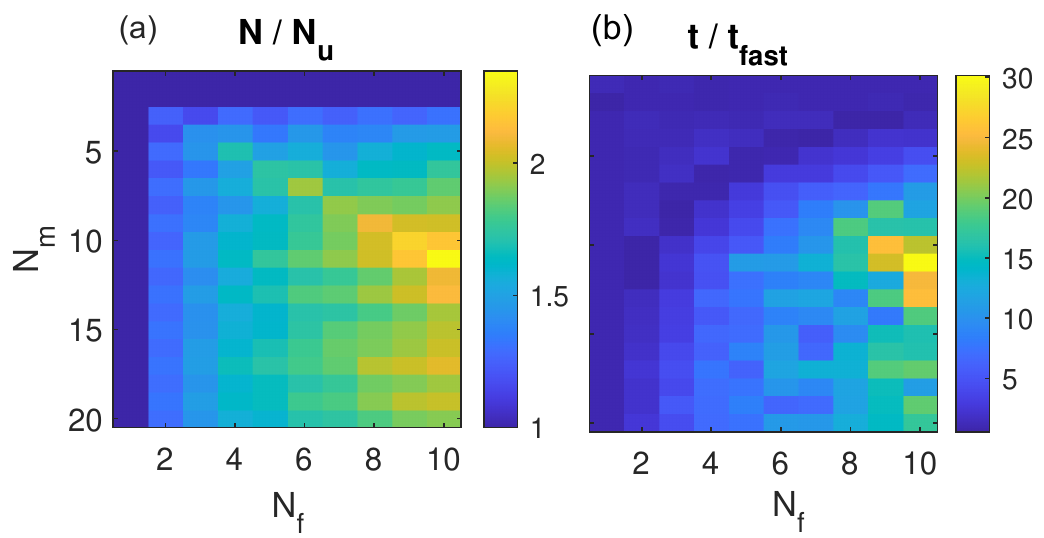}
\caption{(a) $N / N_u$; (b) $t / t_{\textrm{fast}}$, where $t$ and $t_{\textrm{fast}}$ are CPU times for \eqref{ssdp} and the fast program, respectively.}
\label{fast_improve}
\end{figure}

\begin{figure}[t]
\centering
\includegraphics[width=8.5cm]{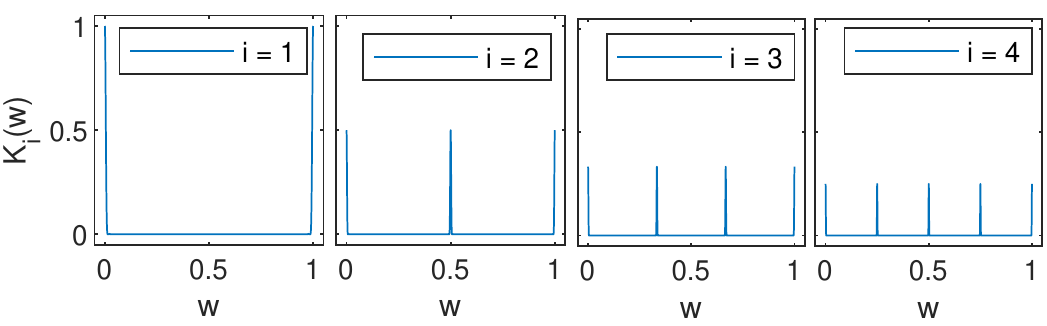}
\caption{Visualization of ${K}_i(w)$ for $i \in \{1, 2, 3, 4\}.$}
\label{fejer_plot}
\end{figure}

\subsection{Fast Algorithm}
\label{fast_ag}
We notice that many rows in the matrix $\mathbf{H}$ are all zero, yet they contribute to the size of the SDP constraint in \eqref{sdp}. This inspires us to come up with a fast algorithm which only includes the non-zero rows of $\mathbf{H}$ in the SDP constraint. This fast algorithm generalizes the method to any frequency set. 

In particular, consider a frequency set $\mathcal{F} = {\{F_1, \dots, F_{N_f}\} \cdot f_0}$ with integers $F_1, \dots, F_{N_f}$ and define $\mathcal{U} = \{m \cdot f |m \in \{0, \dots, N_m - 1\}, f \in \{F_1, \dots, F_{N_f} \} \}$ with cardinality $N_u$. The ratio of $N/N_u$ in Fig. \ref{fast_improve}(a) shows a factor of 2 in savings for large $N_m$ and $N_f$ which gives up to a factor of 30 savings in CPU time (Fig. \ref{fast_improve} (b)). Assume the entries in $\mathcal{U}$ are sorted in ascending order. The matrix $\mathbf{H}_r \in \mathbb{C}^{N_u \times N_f}$ with a reduced number of rows can be expressed in terms of $\mathbf{Q}$ as 
\begin{equation}
\label{H_r}
    \mathbf{H}_r(r, f) = \left\{
\begin{array}{ll}
    \mathbf{Q}(m, f) \quad   &\mbox{for} \;\; (\mathcal{U}_r,f)=(f \cdot (m-1) + 1, f)\\
    0  & \textrm{otherwise};
\end{array}
\right.
\end{equation}
note $r$ is the index of $\mathcal{U}_r = f \cdot (m-1) + 1$. We have the following proposition for an SDP with reduced dimension.  

\begin{proposition}
\label{p1_mbw}
Let $\bm{\psi}(\mathbf{Q}, w)$ be as defined in (\ref{dual_poly_def}). Then $\|\mathbf{Q}\|^{*}_{\mathcal{A}}  \leq 1$ holds if there exists a matrix $\mathbf{P}_{r0} \in \mathbb{C}^{N_u \times N_u} \succeq 0$ such that
\begin{equation}
\label{tr_mbwr}
    \sum_{\substack{i, j\\ \mathcal{U}_j - \mathcal{U}_i = k}} \mathbf{P}_{r0}(i, j) = \delta_k = \left\{
\begin{array}{cc}
\begin{aligned}
  & 1, \quad k = 0,\\
  & 0, \quad k = 1, \dots, N-1
\end{aligned}
\end{array}
\right.
\end{equation}
and such that 
\begin{equation}       
\label{sdp_mbwr}
\left[                 
  \begin{array}{cc}   
    \mathbf{P}_{r0} & \mathbf{H}_r \\
    \mathbf{H}_r^H & \mathbf{I}_{N_f} \\  
  \end{array}
\right]  \succeq 0,             
\end{equation}
where $\mathbf{H}_r$ is defined in \eqref{H_r}.
\end{proposition}
The proof is in the Appendix \ref{proof_fast}. $\hfill\square$

We therefore propose fast alternatives to~\eqref{ssdp} and~\eqref{ssdp_robust} by incorporating the reduced dimension SDP constraint. Note that in Proposition \ref{p1}, we theoretically guaranteed the equivalence between \eqref{ssdp} and \eqref{ANM}. However, we only guarantee the ``if'' part in Proposition \ref{p1_mbw}. Nevertheless, it turns out that the fast algorithm achieves promising performance in the empirical experiments while greatly reducing the computational complexity. The empirical improvement in computational complexity is up to a factor of 30 (see Fig. \ref{fast_improve} (b)). We apply the fast algorithms throughout Sec. \ref{sec:5}.

\subsection{Dual SDP Problem}
Based on \cite[(16)]{yang2017gridless}, we consider the dual problem of \eqref{ssdp}. The dual problem of \eqref{ssdp} is also an SDP, and it can be expressed as

\begin{equation}
\label{ssdp_dual_2}
\begin{aligned}
     &\min_{\mathbf{W}, \mathbf{u}, \mathbf{\widetilde{Y}}} [\mathrm{Tr}(\mathbf{W}) + \mathrm{Tr(Toep}(\mathbf{u}))] \\
    \quad \quad &\textrm{s.t.} 
\left[                 
  \begin{array}{cc}   
    \mathrm{Toep}(\mathbf{u}) & \mathbf{\widetilde{Y}} \\  
    \mathbf{\widetilde{Y}}^H & \mathbf{W} \\  
  \end{array}
\right]  \succeq 0, \mathbf{Y} = \mathcal{R}(\mathbf{\widetilde{Y}}),
\end{aligned}
\end{equation}
where $\mathbf{W} \in \mathbb{C}^{N_f \times N_f}$, $\mathbf{u} \in \mathbb{C}^N$, $\widetilde{\mathbf{Y}} \in \mathbb{C}^{N \times N_f}$, and $\mathrm{Toep}(\mathbf{u})$ is a $N \times N$ Toeplitz matrix with the first column $\mathbf{u}$.

 The derivation of the dual problem is provided in App.~\ref{dual_derive}. After solving~\eqref{ssdp_dual_2}, the DOAs are retrieved using the Vandermonde decomposition of $\mathrm{Toep}(\mathbf{u})$~\cite{yang2017gridless} and the root-MUSIC procedure. Since both \eqref{ssdp} and \eqref{ssdp_dual_2} are strictly feasible, strong duality holds. Therefore, the optimal values for \eqref{ssdp} and \eqref{ssdp_dual_2} must be the same. Note the matrix associated with the PSD constraint for both problems are $(N + N_f) \times (N + N_f)$. We can solve either one of them for DOA estimation.

\section{\label{sec:4}Dual Polynomial Construction}

In Theorem~\ref{thm:cert}, a \textit{sufficient} condition for optimal atomic decomposition was given.  In this section, for certain scenarios, we show that it is possible to construct a dual certificate satisfying \eqref{cert}. This implies the success of the DOA estimation algorithm in the noise-free setting.

Following from \cite{candes2014towards}, we consider an alternative, symmetric index set $\{-2M, ..., 2M\}$ (modified from $\{0, ..., N_m-1\}$), where $M = \frac{N_m - 1}{4}$. Constructing a dual certificate satisfying the requisite properties~\eqref{cert} using the original index set is equivalent to constructing a ``modulated'' dual polynomial $\bm{\psi}(w)$ (note that $\bm{\psi}(w)$ is different from the $\bm{\psi}(\mathbf{Q}, w)$ defined in Sec. III ) on the symmetric index set satisfying 
\begin{equation}
\label{cert_new}
\left\{
\begin{array}{cc}
\begin{aligned}
   & \bm{\psi}(w) = \mathrm{sign}(c_w^*) \overline{\mathbf{x}}_w  \quad \forall w \in \mathcal{W}\\
   & \|\bm{\psi}(w)\|_2 < 1 \quad \forall w \notin \mathcal{W},
\end{aligned}
\end{array}
\right.
\end{equation}
where $\overline{\mathbf{x}}_w(i) := \mathbf{x}_w(i) \cdot e^{-j 2 \pi w i \frac{N_m - 1}{2}}$, 
$\forall i \in \{1, \dots, N_f \}$. Note $|\overline{\mathbf{x}}_w(i)| = |\mathbf{x}_w(i) \cdot e^{-j 2 \pi w i \frac{N_m - 1}{2}}| = |\mathbf{x}_w(i)|$, and $|\bm{\psi}(w)(i) \cdot e^{j 2 \pi w i \frac{N_m - 1}{2}}| = |\bm{\psi}(w)(i)|$. Therefore, as long as $\bm{\psi}(w)$ (associated with the new index set $\{-2M, \dots, 2M\}$) satisfies \eqref{cert_new}, $\overline{\bm{\psi}(w)} := \bm{\psi}(w) \odot [e^{j 2 \pi w \frac{N_m - 1}{2}} \dots e^{j 2 \pi w N_f \frac{N_m - 1}{2}}]^T$ (associated with the original index set) must satisfy \eqref{cert}. Indeed, $\|\bm{\psi}(w)\|_2 = \|\overline{\bm{\psi}(w)}\|_2$ and $\overline{\bm{\psi}(w)} = \mathrm{sign}(c_w^*)\mathbf{x}_w$ for $\forall w \in \mathcal{W}$. In this section, we will construct $\bm{\psi}(w)$ that satisfies \eqref{cert_new}.

In addition, $w \in [0, 1)$ is assumed in this section. Due to the periodicity of the kernel, it is equivalent to consider $w \in [-1/2, 1/2]$ as $w \in [0, 1)$. This assumption indicates that $d = \frac{c}{2f_0}$ needs to be assumed for the following analysis. 


\subsection{Interpolation Kernel}
Inspired by \cite{candes2014towards}, we leverage the $i$-th order squared Fej{\'e}r kernel ${K}_i(w)$ for the dual polynomial construction:
\begin{equation}
\label{fejer}
\begin{aligned}
     {K}_i(w) &:= \frac{1}{iM}\sum_{k=-2M}^{2M} g_M(k)e^{-j2\pi kw \cdot i} \\
     &= \frac{1}{i}\bigg [\frac{\sin(\pi (M+1) wi)}{(M+1) \sin (\pi wi)} \bigg]^4,
\end{aligned}
\end{equation}
where 
\begin{equation}
\label{g_m}
g_M(k) = \frac{1}{M} \sum_{t = \max\{k-M, -M \}}^{\min \{k+M, M\}} \bigg(1 - \frac{|t|}{M} \bigg) \bigg(1 - \frac{|k-t|}{M} \bigg).
\end{equation}
${K}_i(w), i \in \{1, 2, 3, 4 \}$ is shown in Fig. \ref{fejer_plot}. When $i = 1$, ${K}_i(w)$ corresponds to the classical kernel used for the dual polynomial construction in \cite{candes2014towards, tang2013compressed,li2015off, yang2016exact, yang2018sample}. When $i$ increases, the period of the kernel reduces to $1/i$. Therefore, the periodic copies  appears in the visible region $[0, 1)$, and will potentially bring about aliasing for the localization. In addition, note that the amplitude of ${K}_i(w)$ shrinks to $1/i$, which will cancel the scaling factor $i$ of ${K}'_i(w)$. 

We summarize some useful facts for ${K}_i(w)$
\begin{equation}
\label{kernel_property}
    {K}_i(w) = \frac{1}{i} {K}_1(iw) \quad  {K}'_i(w) = {K}'_1(iw) \quad {K}''_i(w) = i  {K}''_1(iw).
\end{equation}

\subsection{Dual Polynomial Construction by Interpolation Kernel}
We construct the dual polynomial vector $\bm{\psi}(w) \in \mathbb{C}^{N_f}$ as follows
\begin{equation}
\label{dual_construct2}
\bm{\psi}(w)\mkern-7mu := \mkern-7mu\left[
\begin{tabular}{@{}c@{}}
 $\sum_{w_k \in \mathcal{W}} [ {\alpha}_{k, 1}{K}_1(w - w_k)+  {\beta}_{k, 1}{K}'_1(w - w_k)]$  \\ \vdots  \\ $\sum_{w_k \in \mathcal{W}} [{\alpha}_{k, N_f}{K}_{N_f}(w\mkern-5mu - \mkern-5muw_k) + {\beta}_{k, N_f}{K}'_{N_f}(w\mkern-5mu - \mkern-5mu w_k) \mkern-1mu]$
\end{tabular}
\right] \mkern-7mu,
\end{equation}
where ${K}'_{i}(w - w_k)$ is the first order derivative for ${K}_{i}(w - w_k)$. 

The constructed dual polynomial in \eqref{dual_construct2} is valid if there exists ${\alpha}_{k,i}$ and ${\beta}_{k, i}$ ($i = 1, ..., N_f$) that satisfy \eqref{cert}. To satisfy (\ref{cert}), for each frequency, we must have \cite{candes2014towards}

\begin{equation}
\label{linear}
\left[
\begin{tabular}{@{}c@{}}
$\textbf{D}_{i, 0}$ \quad $\textbf{D}_{i, 1}$ \\
$\textbf{D}_{i, 1}$ \quad $\textbf{D}_{i, 2}$
\end{tabular}
\right] \left[
\begin{tabular}{@{}c@{}}
${\alpha}_{1i}$ \\ \vdots \\ ${\alpha}_{Ki}$ \\ ${\beta}_{1i}$ \\ \vdots \\ ${\beta}_{Ki}$

\end{tabular}
\right] = \left[\begin{tabular}{@{}c@{}}
$\mathrm{sign}(c_w^{*})\overline{\mathbf{x}}_{w_1}(i)$ \\ \vdots \\ $\mathrm{sign}(c_w^{*})\overline{\mathbf{x}}_{w_K}(i)$ \\ 0 \\ \vdots \\ 0
\end{tabular}
\right] = \left[
\begin{tabular}{@{}c@{}}
$\mathbf{c}_i$ \\ $\mathbf{0}_{K}$ 
\end{tabular}
\right],
\end{equation}
where ($K_i^{(l)}$ is the $l$-th order derivative of $K_i$)
\begin{equation}
\label{D_def}
[\textbf{D}_{i, l}]_{mn} := {K}_i^{(l)}(w_m - w_n), \quad m, n \in \{1, ..., K \}, l \in \{0, 1, 2 \},
\end{equation}
and $\mathbf{c}_i := [\mathrm{sign}(c_w^{*})\overline{\mathbf{x}}_{w_1}(i) \dots \mathrm{sign}(c_w^{*})\overline{\mathbf{x}}_{w_K}(i)]^T \in \mathbb{C}^{K}$. $\bm{\psi}(w)$ can be expressed as 
\begin{equation}
\label{c_def}
    \bm{\psi}(w) = [\sum_{k = 1}^K \mathbf{c}_1(k) \dots \sum_{k = 1}^K \mathbf{c}_{N_f}(k)]^T.
\end{equation}

One sufficient condition to ensure the existence for ${\alpha}_{k,i}$ and ${\beta}_{k, i}$ ($i = \{1, ..., N_f\}$) is that
\begin{equation}
\label{Ki}
\mathbf{K}_i := \left[
\begin{tabular}{@{}c@{}}
$\textbf{D}_{i, 0}$ \quad $\textbf{D}_{i, 1}$ \\
$\textbf{D}_{i, 1}$ \quad $\textbf{D}_{i, 2}$
\end{tabular}
\right] \in \mathbb{C}^{2K \times 2K}  
\end{equation}
is invertible for any $i \in \{1, ..., N_f\}$, which means $\mathrm{rank(\mathbf{K}_i)} = 2K$.  Then, the solution to \eqref{linear} is uniquely determined by inverting $\mathbf{K}_i$. Unfortunately, the invertibility of $\mathbf{K}_i$ may not be guaranteed in general.

\subsection{Single Source Analysis}
We begin with single source analysis ($K = 1$). For one source, there is no separation condition or risk of collision to consider in the analysis. The constructed $K_i(w)$ is guaranteed to satisfy \eqref{cert} as stated in the theorem.

\begin{theorem}
Suppose $K = 1$ (DOA is $w_1$), and  $\overline{\mathbf{x}}_{w_1}(i) \neq 0$ for $\forall i \in \{1, ..., N_f\}$. We then have 
\begin{equation}
\label{cert_1source}
\left\{
\begin{array}{cc}
\begin{aligned}
   & \bm{\psi}(w) = \mathrm{sign}(c_w^*) \overline{\mathbf{x}}_w  \quad w = w_1\\
   & \| \bm{\psi}(w)\|_2 < 1 \quad \forall w \neq w_1.
\end{aligned}
\end{array}
\right.
\end{equation}
\end{theorem}
\textit{Proof}  Since $K = 1$, (\ref{linear}) reduces to 
\begin{equation}
\begin{aligned}
\left[
\begin{tabular}{@{}c@{}}
${K}_i(0)$ \quad ${K}'_i(0)$ \\
${K}'_i(0)$ \quad ${K}''_i(0)$ 
\end{tabular}
\right] \left[
\begin{tabular}{@{}c@{}}
${\alpha}_{1i}$ \\ ${\beta}_{1i}$
\end{tabular}
\right] &= \left[
\begin{tabular}{@{}c@{}}
$1/i$ \quad 0 \\
0 \quad ${K}''_i(0)$ 
\end{tabular}
\right] \left[
\begin{tabular}{@{}c@{}}
${\alpha}_{1i}$ \\ ${\beta}_{1i}$
\end{tabular}
\right] \\
&= \left[\begin{tabular}{@{}c@{}}
$\mathrm{sign}(c_w^*)\overline{\mathbf{x}}_{w_1}(i)$ \\ 0
\end{tabular}
\right].   
\end{aligned}
\end{equation}
Hence ${\alpha}_{1i} = i \cdot \mathrm{sign}(c_w^*)\overline{\mathbf{x}}_{w_1}(i)$ and ${\beta}_{1i} = 0$. Furthermore, 
\begin{equation}
\|\bm{\psi}(w)\|_2^2 = \sum_{i = 1}^{N_f} |{\alpha}_{1i}{K}_i(w - w_1)|^2.
\end{equation}

When $w = w_1$, $\bm{\psi}(w) = [\alpha_{11}K_1(0) \dots \alpha_{1N_f}K_{N_f}(0)]^T = \mathrm{sign}(c_w^*)[\overline{\mathbf{x}}_{w_1}(1) \dots \overline{\mathbf{x}}_{w_1}(N_f)]^T = \mathrm{sign}(c_w^*) \overline{\mathbf{x}}_{w}$ and $\|\bm{\psi}(w)\|_2^2 =\|\overline{\mathbf{x}}_{w_1}\|_2^2 = 1.$

For $w \neq w_1$, suppose ${\alpha}_{1i} = i \cdot \mathrm{sign}(c_w^*) \overline{\mathbf{x}}_{w_1}(i)  \neq 0$ for $\forall i \in \{1, ..., N_f\}$, and notice that ${K}_i(w-w_1) < {K}_i(0) = 1/i$. Therefore
\begin{equation}
\|\bm{\psi}(w)\|_2^2 = \sum_{i = 1}^{N_f} |{\alpha}_{1i}{K}_i(w - w_1)|^2 < \sum_{i = 1}^{N_f} |{\alpha}_{1i}{K}_i(0)|^2 = 1.    
\end{equation}
Therefore, \eqref{cert_1source} must hold. $\hfill\square$


\subsection{Multiple Source Analysis}
\label{two_source}
The analysis is now extended to multiple source cases. For the existing ANM based methods, if there is more than one source, a minimum separation condition is assumed \cite{candes2014towards, tang2013compressed, li2015off, yang2016exact}. However, in our signal model, we have to consider the potential for aliasing and collisions (see Sec.\ \ref{aliase_collision}).

We first define the separation of $\mathcal{W}$ for the $i$-th frequency $\Delta(\mathcal{W}^i)$ as the closest wrap-around distance between two \textit{distinct} DOAs $w_m, w_n$
\begin{equation}
\begin{aligned}
     \Delta(\mathcal{W}^i) := \inf_{w_m, w_n \in \mathcal{W}} & \min \{ i|w_m - w_n|\mod 1, \\
     & \qquad 1 - (i|w_m - w_n|\mod 1) \}.
\end{aligned}
\end{equation}
Note that although $|w_m - w_n| \in [0, 1)$, for $i \geq 2$, $i|w_m - w_n|$  can be greater than $1$. Due to the periodicity of the interpolation kernel, we keep only the fractional part of $i|w_m - w_n|$ in the definition of the separation. 
We first introduce the concepts of aliasing and collision before our analysis. 

\subsubsection{Aliasing and Collision}
\label{aliase_collision}

\textit{Aliasing.} Because of the wrap-around nature of $\mathbf{a}(i, w)$, when $d > \frac{\lambda_{N_f}}{2}$ there will be aliasing peaks in the higher frequencies.  Aliasing can happen even for the single source case. Specifically, based on \cite{van2004optimum}, if the temporal frequency $f \cdot f_0 $ satisfies 
\begin{equation}
    f \cdot f_0 \geq \frac{c}{d} \frac{1}{1 + |\cos (\theta)|},
\end{equation}
then aliasing peaks enter into the visible region $[-1/2, 1/2]$ and that frequency experiences \textit{aliasing}. When $d = c/(2f_0)$ and $\theta \in [0^\circ, 180^\circ]$, aliasing happens for all $f \geq 2$. In addition to the peak associated with the ground-truth DOA $w$, there are aliasing peaks with DOAs $\bar{w} = w \pm \frac{k}{f}, (k < f, k \in \mathbb{N}_+)$. It can be shown that 
\begin{equation}
    \mathbf{a}(f, w) = \mathbf{a}(f, \bar{w}).
\end{equation}

Aliasing happens for the single frequency beamforming \cite{van2004optimum} provided that the temporal frequency is high enough. In \cite{nannuru2019sparse}, the authors demonstrate that multiple frequencies can overcome aliasing for conventional beamforming (CBF) and sparse Bayesian learning (SBL) methods. 

\textit{Collision.} 
One consequence of aliasing is the possibility of \textit{collision} of multiple DOAs. Collision occurs when one DOA  lies exactly in the positions of the aliasing peaks of another source. Formally, suppose there are $K = 2$ distinct DOAs $w_1$ and $w_2$ ($w_1, w_2 \in [-1/2, 1/2]$). $w_1$ and $w_2$ are said to have \textit{collision} in the $i$-th frequency if 
\begin{equation}
\label{aliasing_manifold}
    \mathbf{a}(i, w_1) = \mathbf{a}(i, w_2).
\end{equation}
Such collision occurs whenever $w_1$ and $w_2$ satisfy
\begin{equation}
\label{aliase_sep}
    |w_1 - w_2| = \frac{k}{i} \quad (i \in \{2, \dots, N_f\}, k < i, k \in \mathbb{N}_+).
\end{equation}
When collision occurs in the $i$-th frequency bin, it is verified that the $pi$-th ($p \geq 2, p \in \mathbb{N}_+$) frequency bins also have collision.


For CBF and SBL, collision may bring about ambiguities in the source power (and amplitude) estimation as these two sources share the same array manifold vector.

\begin{figure}[t]
\centering
\includegraphics[width=8.5cm, height = 10.3 cm]{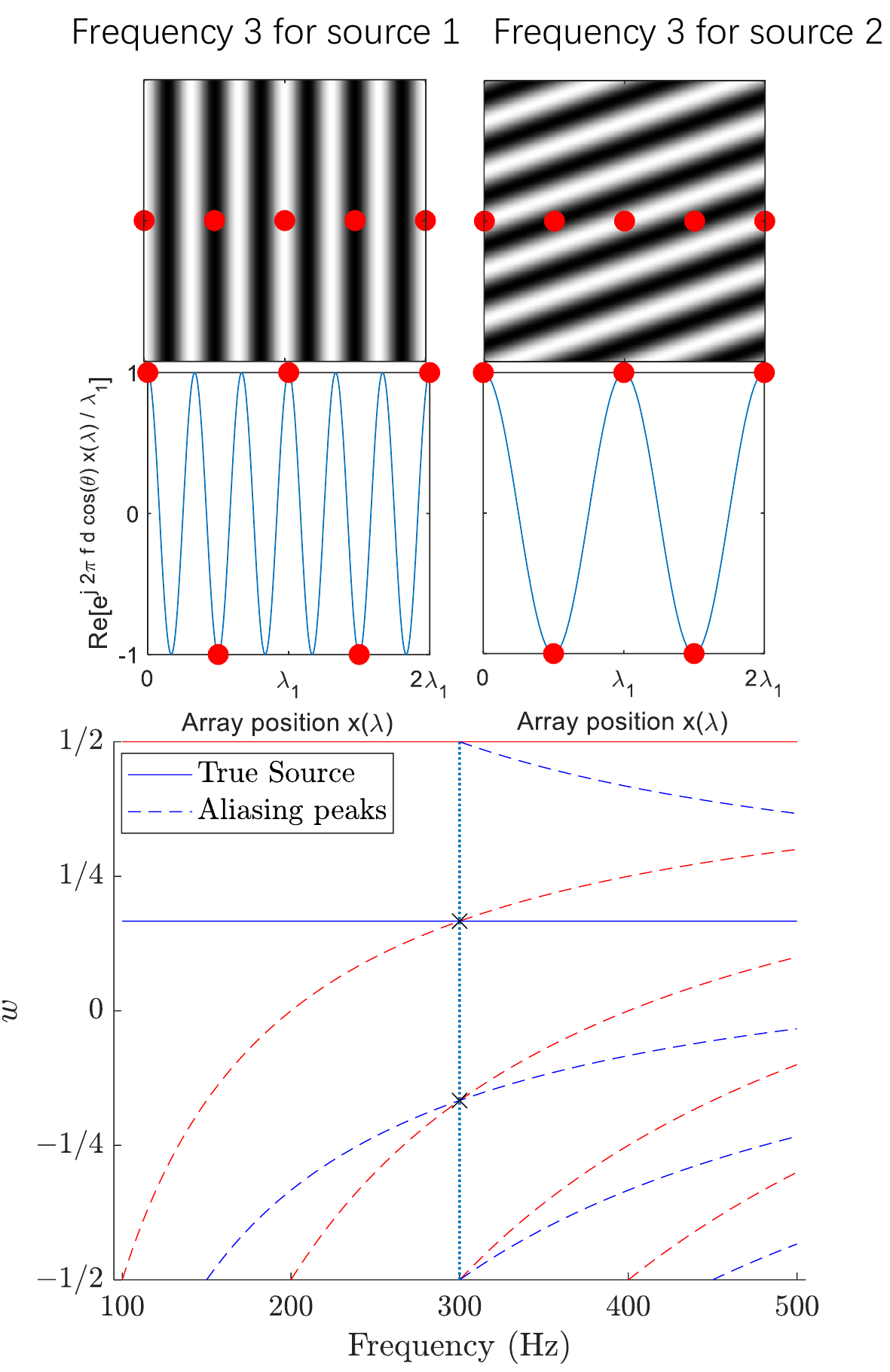}
\caption{Collision demonstration. $K = 2, N_f = 5, w_1 = 1/2, w_2 = 1/6$. (a--b) are the same as Fig. \ref{tf_sf}. (c) Red lines indicate $w_1$ and blue lines indicate $w_2$ for the true sources (solid), and the aliased signal (dashed). Collision occurs at 300 Hz. }
\label{Collision_demo}
\end{figure}

As an example, let $N_f = 5$, $f_0 = 100$ Hz, $w_1 = 1/2$, and $w_2 = 1/6$. Then $\mathbf{a}(3, w_1) = \mathbf{a}(3, w_2)$ and so these two sources collide in the third frequency bin. As Fig.~\ref{Collision_demo} (a)-(b) illustrate, the spatial samples obtained from all sensors are the same at that frequency. In addition, collision can  be interpreted as the intersection of the true DOA of one source and the aliasing peaks of another source. In Fig.~\ref{Collision_demo} (c), it is clear the collision exists in the third frequency (300 Hz).

For our ANM problem, if \eqref{aliase_sep} is satisfied, based on \eqref{aliasing_manifold}, we must have 
\begin{equation}
\label{dual_cons}
\mathbf{q}_{i}^{H}\mathbf{a}(i, w_1) = \mathbf{q}_{i}^{H}\mathbf{a}(i, w_2).
\end{equation}
Based on the definition of the dual polynomial in \eqref{dual_poly_def}, the $i$-th entry of $\bm{\psi}(\mathbf{Q}, w_1)$ and $\bm{\psi}(\mathbf{Q}, w_2)$ must therefore be equal. This serves as an additional constraint for the dual polynomial.  We refer to \eqref{aliase_sep} as the \textit{exact collision} case. Collisions complicate the construction of a dual polynomial that satisfies the optimality condition (see~\eqref{dual_cons}). However, we observe that in the numerical experiments, the method still works in the presence of exact collisions (See Fig. \ref{Case1_Nf_5} (a)).

\subsubsection{Case Classification}
With multiple sources, depending on the true DOAs, we have three possible cases:
\begin{itemize}
    \item Case 1: There exists an \textit{exact collision}. An exact collision in the $i$-th frequency is defined as
    \begin{equation}
    \label{collision}
        |w_m - w_n| = \frac{k}{i} \quad (i \in \{1, \dots, N_f\}, k < i).
    \end{equation}
    for some DOAs $w_m,w_n$. For example, suppose $w_1 = 1/2$, $w_2 = 1/6$, $N_f = 6$. Since $|w_1 - w_2| = 1/3$, the third frequency has collision. Indeed, as shown in Fig.~\ref{Collision_demo}, the spatial samples obtained from all sensors are the same in the third frequency. Notice also that $|w_1 - w_2| = 2/6 = 1/3$, so the sixth frequency also has collision.
    \item Case 2: There exists a \textit{near collision}. A near collision in the $i$-th frequency is defined as 
    \begin{equation}
    \label{near_collision}
        |w_m - w_n| = \frac{k}{i} \pm \epsilon \quad (i \in \{1, \dots, N_f\}, k < i),
    \end{equation}
    for some $w_m,w_n$ for sufficiently small $\epsilon > 0$. The upper bound of $\epsilon$ is proportional to $1/N_m$. For example, suppose $w_1 = 1/4$, $w_2 = 0.001$, $N_f = 6$, and the minimum separation condition $\Delta_{\min} = 0.01$. Then $|w_1 - w_2| = 1/4 - 0.001 = 1/4 - \epsilon$ with $\epsilon = 0.001 < \Delta_{\min}$. Therefore, the fourth frequency has a near collision. 
    \item Case 3: There are no collisions or near collisions across all $N_f$ frequencies. For example, suppose $w_1 = 1/4$, $w_2 = 1/10$, $N_f = 6$, and $\Delta_{\min} = 0.01$. It can be easily shown that there is no collision or near collision for any $i \in \{1, \dots, N_f\}$.
\end{itemize}

\subsubsection{Case 1 and 2 Study}
For Case 1 and 2, an analytical guarantee is hard to obtain due to the singularity of $\mathbf{K}_i$. We list some properties for Case 1 in the Appendix \ref{appendix_collision}. Although an analytical guarantee is hard to obtain, we find the method~\eqref{ssdp} can perform well in Case 1 (See Fig.~\ref{Case1_Nf_5} (a)). However, directly solving \eqref{ssdp} for Case 2 does not give a satisfactory performance (See Fig.~\ref{Case2_dual}).  To resolve the near collision issue in \eqref{ssdp_robust}, we proposed a robust solution in \eqref{ssdp_robust}.  The robust solution applies $\ell_{1, 2}$ regularization to nullify the contribution from the near collision frequencies. The numerical examples (see Fig.~\ref{Case2_dual}) demonstrate the effectiveness of the $\ell_{1, 2}$ regularization in suppressing near collisions. 

\subsubsection{Case 3 Analysis}
For Case 3, there is no collision and therefore the theoretical analysis becomes tractable. Under a uniform amplitude assumption, we draw  analytical conclusions on $\|\bm{\psi}(w)\|_2$ in Theorem \ref{case2_theorem}.  
\begin{theorem}
\label{case2_theorem}
If \textcolor{black}{the amplitude is uniform across frequencies for each source (i.e. $|\mathbf{x}_{w}(1)| = \dots = |\mathbf{x}_{w}(N_f)| = 1/\sqrt{N_f}$ for all $w \in \mathcal{W}$)}, $\Delta(\mathcal{W}^i) \geq 4/(N_m - 1)$ and $N_m \geq 257$, then $\|\bm{\psi}(w)\|_2 < 1$ for $w \notin \mathcal{W}$.
\end{theorem}
\textcolor{black}{\textbf{Remark} The assumptions on the uniform amplitudes and the number of sensors are made to facilitate the proof and may not be necessary in practice. Intuitively, the uniform amplitude assumption prevents  certain frequency bins from dominating the source amplitudes, which in the extreme case could transform the multi-frequency model into the single-frequency model. The assumption on the number of sensors is used to bound the Fej{\'e}r kernel. Note also that the separation assumption implicitly implies an upper bound for the source number $K$. \\
\textit{Proof} See Appendix \ref{appendix_proof} and the following paragraphs.}

With the first $K$ constraints in~\eqref{linear}, the constructed $\bm{\psi}(w)$ automatically satisfies the first equality condition in (\ref{cert}) \textcolor{black}{as $\bm{\psi}(w)$ satisfies \eqref{c_def}}. However, we also need to show that with the last $K$ equality constraints in \eqref{linear}, the constructed $\bm{\psi}(w)$ satisfies the second inequality condition in (\ref{cert}) (i.e. $\|\bm{\psi}(w)\|_2 < 1$), and we prove Theorem \ref{case2_theorem} to guarantee that. Inspired by \cite{candes2014towards}, to bound $\|\bm{\psi}(w)\|_2$, ${\alpha}$ and ${\beta}$ in (\ref{linear}) need to be bounded first. To simplify the derivation, we prove the case when $K = 2$ in the following sections. The result can be generalized to $K > 2$ with the same reasoning. 

Supposing that $K = 2$, (\ref{linear}) is simplified as a 4 $\times$ 4 system of equations. Note that $i = 1$ is the classical case \cite{candes2014towards, tang2013compressed}. Since collision is absent in this case, the matrix $\mathbf{K}_i$ defined in \eqref{Ki} is invertible (\textcolor{black}{for detailed reasoning, see Appendix \ref{invert_K}}). Therefore, the solution for \eqref{linear} is \textit{uniquely} identified as


\begin{equation}
\begin{aligned}
\left[
\begin{tabular}{@{}c@{}}
${\alpha}_{1i}$ \\ ${\alpha}_{2i}$ \\ ${\beta}_{1i}$ \\ ${\beta}_{2i}$

\end{tabular}
\right] &= \left[\begin{tabular}{@{}c@{}}
$\mathbf{D}_{i, 0}$ \quad $\mathbf{D}_{i, 1}$ \\
$\mathbf{D}_{i, 1}$ \quad $\mathbf{D}_{i, 2}$ \end{tabular}
\right]^{-1}   \left[\begin{tabular}{@{}c@{}}
$\mathrm{sign}(c_w^{*})\overline{\mathbf{x}}_{w_1}(i)$ \\ $\mathrm{sign}(c_w^{*})\overline{\mathbf{x}}_{w_2}(i)$ \\ 0 \\ 0
\end{tabular}
\right] \\ 
&= \left[\begin{tabular}{@{}c@{}}
$\mathbf{S}_i^{-1}$  \\
$-\mathbf{D}_{i, 2}^{-1}\mathbf{D}_{i, 1}\mathbf{S}_i^{-1}$  \end{tabular}
\right]\left[\begin{tabular}{@{}c@{}}
$\mathrm{sign}(c_w^{*})\overline{\mathbf{x}}_{w_1}(i)$ \\ $\mathrm{sign}(c_w^{*})\overline{\mathbf{x}}_{w_2}(i)$\end{tabular}
\right].
\end{aligned}
\end{equation}
where the Schur complement $\mathbf{S}_i := \mathbf{D}_{i, 0} - \mathbf{D}_{i, 1}\mathbf{D}_{i, 2}^{-1}\mathbf{D}_{i, 1}$.

Define ${\mathbf{\alpha}}_i := [{\alpha}_{1i} \quad {\alpha}_{2i}]^T$ and ${\mathbf{\beta}}_i := [{\beta}_{1i} \quad {\beta}_{2i}]^T$. The following lemma gives upper bounds for $\|{\alpha}_i\|_{\infty}$ and $\|{\beta}_i\|_{\infty}$.

\begin{lemma}
\label{bound_alpha_beta}
If $\Delta(\mathcal{W}^i) \geq 4/(N_m - 1) = 1/M$ and $N_m \geq 257$ (or \textcolor{black}{$f_c := 2M \geq 128$}), then 
\begin{equation}
\textrm{(1)} \|{\alpha}_i\|_{\infty} \leq i \cdot 1.008824 ~\text{and}~ \|{\beta}_i\|_{\infty} \leq \frac{3.294 \times 10^{-2}}{f_c}.
\end{equation}
(2) If the amplitude is uniform across frequencies for  
each source (i.e. $|\mathbf{x}_{w}(1)| = \dots = |\mathbf{x}_{w}(N_f)| = 1/\sqrt{N_f}$ for all $w \in \{w_1, w_2\}$), we further have
\begin{equation}
    \|{\alpha}_i\|_{\infty} \leq \frac{i \cdot 1.008824}{\sqrt{N_f}} ,  \|{\beta}_i\|_{\infty} \leq \frac{3.294 \times 10^{-2}}{f_c\sqrt{N_f}}.
\end{equation}
\end{lemma}
\textit{Proof} See Appendix \ref{bound_lemma} for (1). The proof for (2) is similar to that of Lemma 4.3 with the additional condition $\bigg\|\left[\begin{tabular}{@{}c@{}}
 $\mathrm{sign}(c_w^{*})\mathbf{x}_{w_1}(i)$ \\ $\mathrm{sign}(c_w^{*})\mathbf{x}_{w_{2}}(i)$\end{tabular}
\right]\bigg\|_{\infty} = \frac{1}{\sqrt{N_f}}$. $\hfill\square$


Now that the upper bounds for $\|{\alpha}_i\|_{\infty}$ and $\|{\beta}_i\|_{\infty}$ have been obtained, $\|\bm{\psi}(w)\|_2$ can be further bounded. The remaining steps for bounding $\|\bm{\psi}(w)\|_2$ are available in Appendix \ref{appendix_proof}.

\section{Numerical Results}
\label{sec:5}
\subsection{Case Studies}
\label{case_eg}
We evaluate our method for the 3 cases mentioned in Sec. \ref{two_source}. The noisy case is also evaluated. 

The simulation setup for the following examples
is $K$ incoherent sources have DOAs $\theta =  \{\theta_1, \dots \theta_K\}$ ($90^\circ$ is considered broadside). Assume $c = 340$ m/s, $f_0 = 100$ Hz, a uniform linear array with $N_m $ sensors and spacing $\mtight d = \frac{c}{2f_0}$. The temporal frequencies of the sources are $\{1, \dots, N_f\} \cdot f_0$ Hz. The amplitude vectors $\mathbf{x}_w$ of the 3 sources are randomly generated with standard complex normal distribution $\mathcal{CN}(0, 1)$ and  then normalized  so that $\|\mathbf{x}_w\|_2 = 1$. In Fig. \ref{Case1_Nf_5}, $100$ realizations are evaluated and in each realization, $\mathbf{x}_w$ will be different. We plot the distribution of the DOA estimation of these realizations in the histogram. All $c_w = 1$. The noise  for each frequency $\mathbf{w}_f$ is randomly generated from the distribution $\mathcal{CN}(0, \sigma^2)$ and then scaled to fit the desired signal-to-noise ratio (SNR) defined as 
\begin{equation}
    \mathrm{SNR} = 20 \log_{10} \frac{\|\mathbf{X}\|_F}{\|\mathbf{W}\|_F}.
\end{equation}
This setup is applied in all of the examples in the Sec.\ \ref{case_eg} unless otherwise specified.

\subsubsection{The Dual Polynomial for Case 2}
\begin{figure}[!t]
\centering
\includegraphics[width=8.5cm]{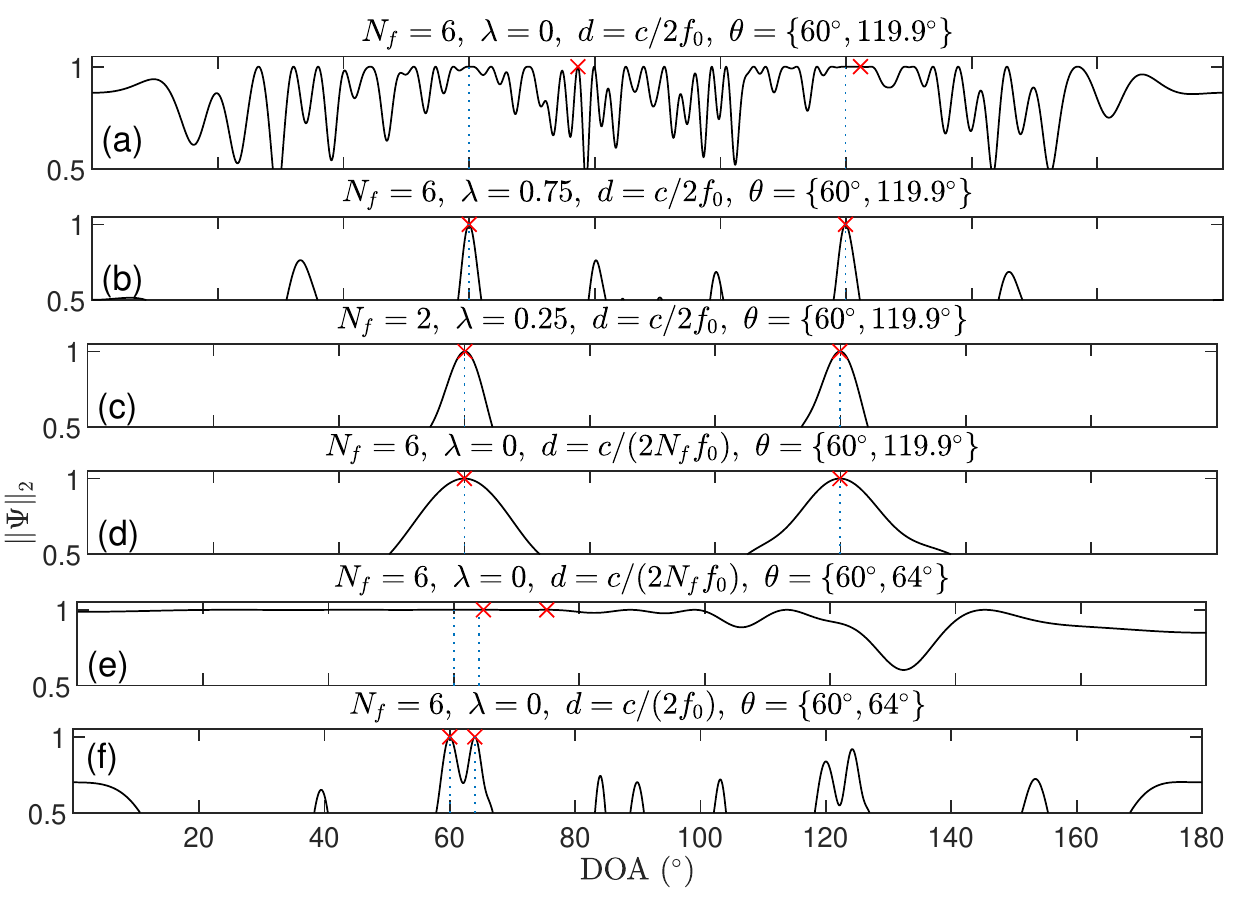}
\caption{$\|\bm{\psi}(\mathbf{Q}, w)\|_2$ versus DOA $\theta$ for Case 2. $N_m = 12$, $f_0 = 100$ Hz, $d = \frac{c}{2f_0}$, and $\mathbf{x}_w \sim \mathcal{CN}(0, 1)$, $K = 2$. ``$\times$'' indicates the peak, and the dashed lines indicate the ground-truth DOAs.}
\label{Case2_dual}
\end{figure}

For Case 2, if $d = \frac{c}{2f_0}$, then all of the frequencies other than the first frequency will have the risk of near collision. To overcome this issue, robust ANM (see \eqref{ssdp_robust}) needs to be employed to suppress the near collision. An alternative way to suppress the collision is to choose a smaller spacing $d = \frac{c}{2N_f f_0}$ so that the collision can be completely avoided for all frequencies. These two collision suppression methods will be examined. Suppose there are $K = 2$ incoherent sources. In this case, if $N_f \geq 2$, then the $2n$-th ($n$ is any positive integer) frequency will have the near collision.  The dual polynomial for different $N_f$, $\lambda$, $d$, and $\theta$ ($\lambda$ is the regularization hyper-parameter in (\ref{ssdp_robust})) can be seen in Fig. \ref{Case2_dual}. For the regularization parameter $\lambda$, we empirically choose it proportional to $N_f$ (i.e. $\lambda = k \cdot N_f$, with $k = 0.125$ in particular for Fig. \ref{Case2_dual}). The intuition behind is that for more frequencies, the near collision is more likely to happen. However, since the regularization can bring bias, a smaller $\lambda$ is more favorable in practice.

From Fig. \ref{Case2_dual} (a), if we only solve (\ref{ssdp}) without regularization, numerous spurious peaks are an obstacle for identifying source positions. However, with  regularization, the  dual polynomial peaks become precise indicators for the source positions (See Fig. \ref{Case2_dual}(b--c)). When $N_f = 6$, the near collision frequencies are the 2nd, 4th, and 6th frequencies. Fig.\ \ref{Case2_dual} (d) demonstrates the success of choosing a smaller spacing $d = \frac{c}{2N_ff_0}$ in collision suppression without regularization. However, there are potential limitations for smaller spacing. Comparing Fig. \ref{Case2_dual} (e) and (f), the smaller aperture cannot resolve the close sources while the larger aperture can. Thus, although the smaller aperture can avoid the collision, it has lower spatial resolution. We leave the theoretical analysis for choosing the regularization hyperparameter $\lambda$ as future work.

\subsubsection{Case 1, 3, and Noisy cases}
\begin{figure}[!t]
\centering
\includegraphics[width=8.5cm, height = 6cm]{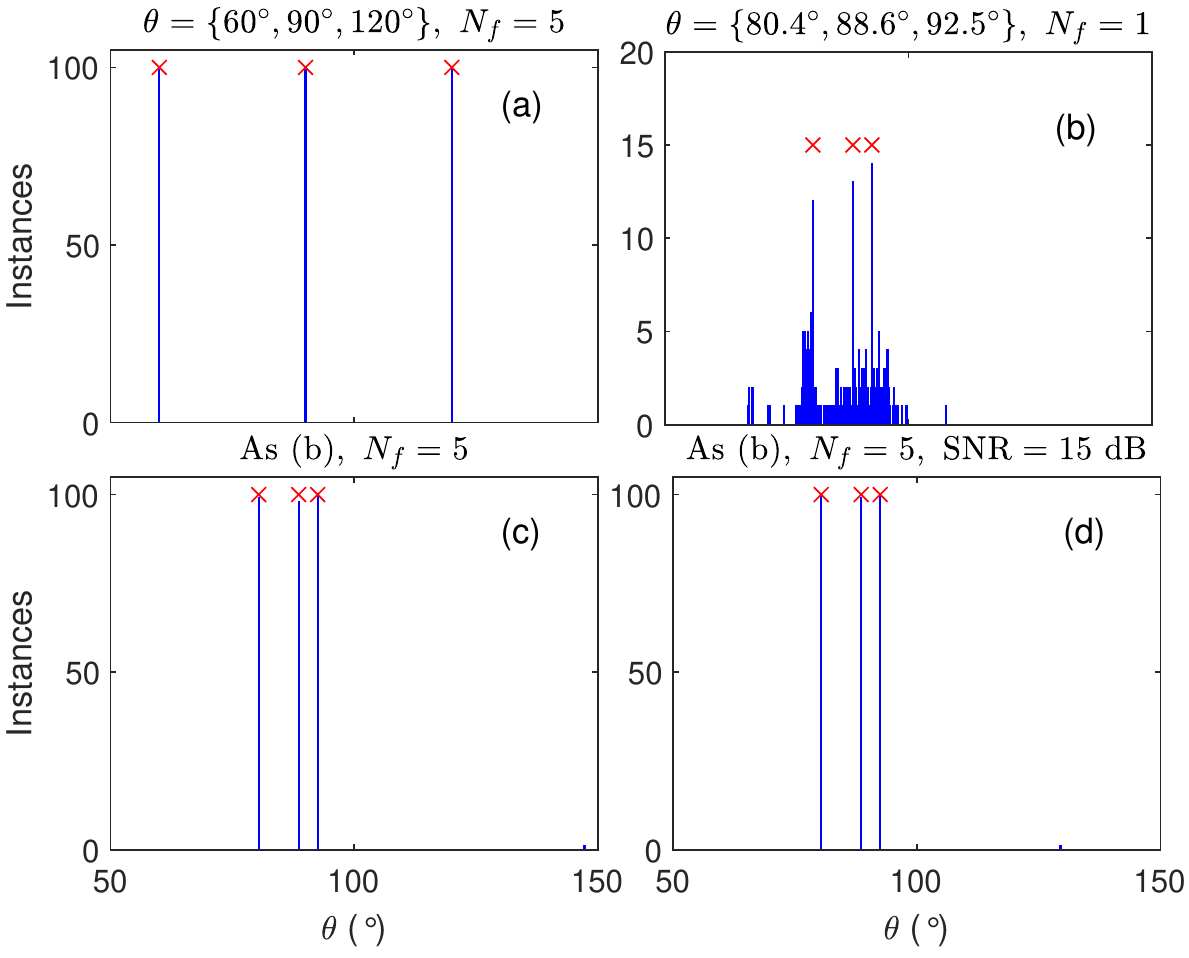}
\caption{Histogram of the estimated DOA $\hat{\theta}$ for 100 realizations with true DOAs ($\times$). $N_m = 12$, $f_0 = 100$ Hz, $d = \frac{c}{2f_0}$, and $\mathbf{x}_w \sim \mathcal{CN}(0, 1)$, $K = 3$. For each realization, $\mathbf{x}_w$ will be different. No noise is present except for (d) where SNR is 15 dB.}
\label{Case1_Nf_5}
\end{figure}

The histograms for these cases are plotted in Fig. \ref{Case1_Nf_5}.  Since $|w_1 - w_2| = |w_2 - w_3| = 0.25$ and $|w_1 - w_3| = 0.5$, there are collisions in both the second and fourth frequencies. From Fig. \ref{Case1_Nf_5} (a), all of the instances in the histogram are nevertheless concentrated in the ground-truth positions, which shows the proposed method can capture the ground-truth positions accurately and has the robustness to the exact collisions. The robustness to the exact collisions is attributed to the combination of multiple frequencies. For the collision frequencies, these two sources are essentially one source since they share the same array manifolds for these frequencies (see \eqref{aliasing_manifold}) and they are mixed coherently, which makes it difficult to separate them. For the non-collision frequencies, the two sources are well-separated. Therefore, if we combine all $N_f$ frequencies, the two peaks associated with the DOAs still stand out as long as there exists non-collision frequencies. To demonstrate Case 3,  we compare the single-frequency ($N_f = 1$, see Fig. \ref{Case1_Nf_5} (b)) and multi-frequency ($N_f = 5$, see Fig. \ref{Case1_Nf_5} (c)) scenarios. When $N_f = 1 $, there are many bins that lie in the undesired positions. In contrast, when $N_f = 5$, the bins are mostly concentrated in the ground-truth positions. This example demonstrates the potential benefits of multi-frequency ANM. In Fig. \ref{Case1_Nf_5} (d), the setup is identical to that in Fig. \ref{Case1_Nf_5} (c) except the noise is present. For the noisy case, the empirical value of $\eta$ is chosen as \cite{chi2016guaranteed}
\begin{equation*}
    \eta = \sigma/2 \cdot \sqrt{N_m N_f + 2\sqrt{N_m N_f}}.
\end{equation*}
From Fig. \ref{Case1_Nf_5} (d), the proposed method captures the source positions accurately in the noisy cases.

\subsection{DOA Estimation Performance Evaluation}

To comprehensively evaluate the performance of the proposed method, we conduct Monte Carlo experiments. In all of the experiments in this section, each point represents $MC = 100$  trials, and the root mean square error (RMSE) and mean absolute error (MAE) are computed as 
\begin{equation}
    \mathrm{RMSE} = \sqrt{\frac{1}{MC} \sum_{m = 1}^{MC}\bigg[\frac{1}{K}\sum_{k = 1}^{K}(\hat{\theta}_{mk} - \theta_{mk})^2 \bigg]}.
\end{equation}

\begin{equation}
    \mathrm{MAE} = \frac{1}{MC} \sum_{m = 1}^{MC}\bigg(\frac{1}{K}\sum_{k = 1}^{K}|\hat{\theta}_{mk} - \theta_{mk}|\bigg),
\end{equation}
where $\hat{\theta}_{mk}$, and $\theta_{mk}$ are (sorted) estimated DOAs, and (sorted) ground-truth DOAs for the $k$-th source and $m$-th trial. A maximum threshold of $10^\circ$ was used to penalize the incorrect DOA estimates (see below). $c, f_0, d,$ and the temporal frequencies are the same as those in Sec. \ref{case_eg}. We also compare the proposed method (ANM) with the multi-frequency sparse Bayesian learning (SBL) \cite{nannuru2019sparse} and Cram{\'e}r-Rao bound (CRB) \cite[Eq. (119)]{liang2020review}. For SBL, the spatial angle is discretized into grids with $0.5^\circ$ between the adjacent grid points. Although there are many DOA estimation methods, very few of them have been developed for the multiple-frequency model. Therefore, only SBL and CRB are included for reference. 

\subsubsection{DOA Estimation under Varying SNR} 
\begin{figure}
    \centering
    \includegraphics[width=8.9cm]{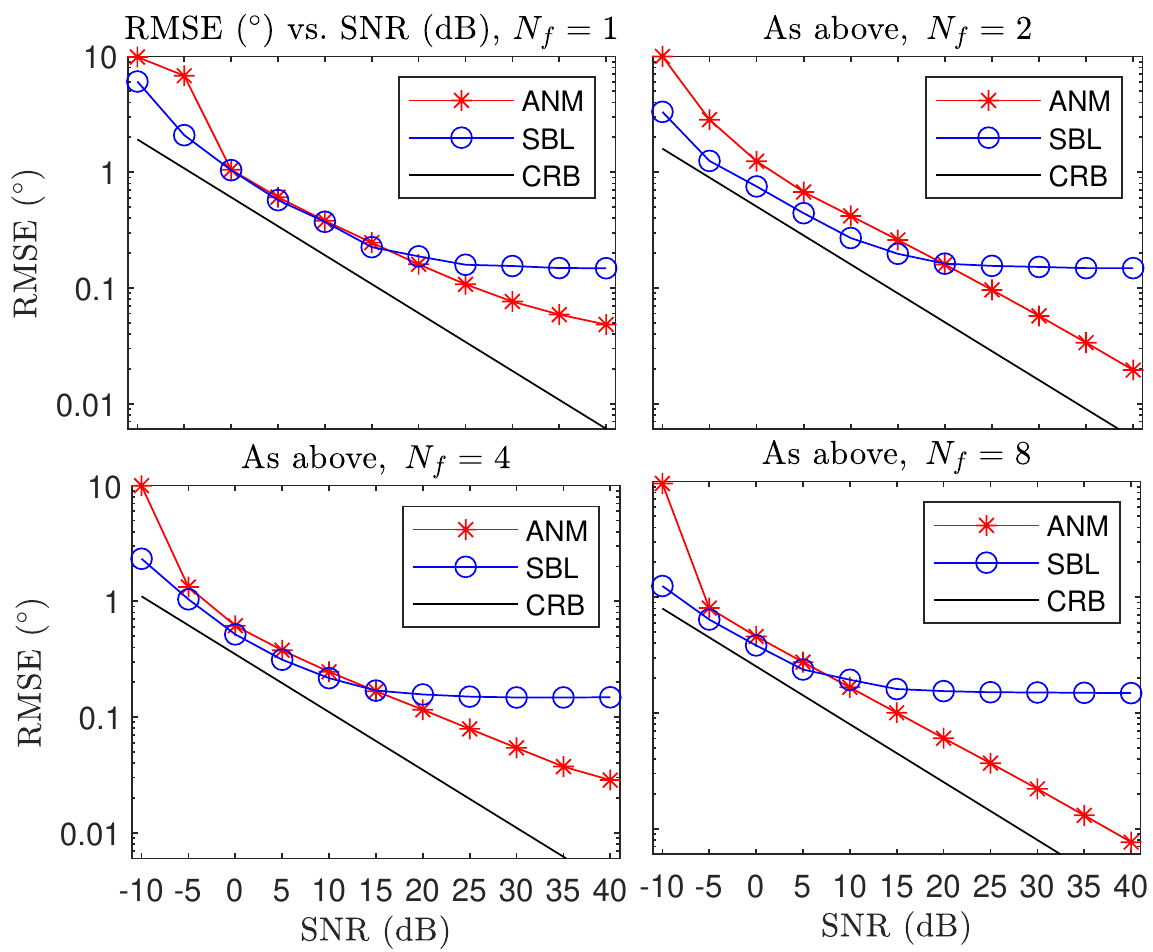}
    \caption{RMSE ($^\circ$) vs. SNR for $d = \frac{\lambda_{N_f}}{2} = \frac{c}{2N_f f_0}$. $N_m = 15$, $K = 3$, $f_0 = 100$ Hz, and the frequency set is $\{1, \dots, N_f\}\cdot f_0$ Hz. The $\ell_{1, 2}$ regularization parameter $\lambda = 0$ for all plots. Each point represents 100 trials. The DOAs for each trial are randomly generated between $[10^\circ, 170^\circ]$ with a minimum angular separation $4/N_m$. $\mathbf{x}_w \sim \mathcal{CN}(0, 1)$.}
    \label{rmse_plot_small}
\end{figure}

We first examine the robustness of ANM to noise. The performance of each algorithm under $d = \frac{\lambda_{N_f}}{2}$ is detailed in Fig. \ref{rmse_plot_small}. Notice that in this setup, there will be no aliasing or collision. Therefore, we can turn off the $\ell_{1,2}$ regularization in \eqref{ssdp_robust}. The proposed algorith outperforms SBL in the high SNR cases. At low SNRs, SBL achieves a better performance since it can estimate the noise power. Note for the SBL with limited $0.5^\circ$ separation, the achievable accuracy for RMSE is $0.125^\circ$. In addition, it turns out that SBL has no failure trials ($\mathrm{RMSE} > 10^\circ$ is defined as failure) starting from SNR $ = 0, -5, -5,$ and $-10$ dB for $N_f = 1, 2, 4,$ and $8$. For ANM, the same happens for SNR $ = 0, 0, -5,$ and $-5$ dB. Therefore, for both SBL and ANM, the performance improves in the low SNR region, which demonstrates the enhanced robustness to noise for the multi-frequency processing. 

\begin{figure}
    \centering
    \includegraphics[width=8.9cm]{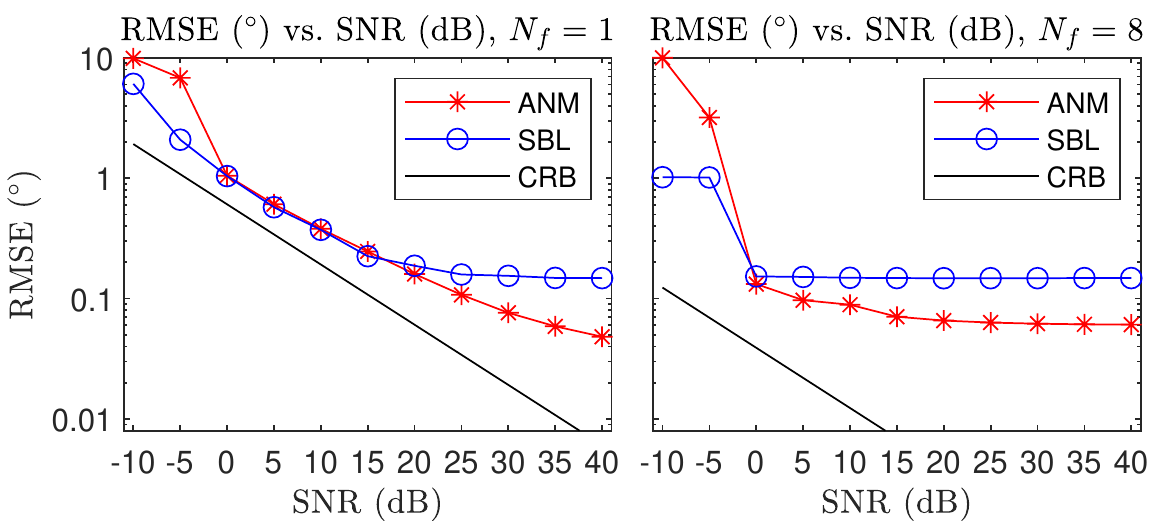}
    \caption{RMSE ($^\circ$) vs. SNR for $d = \frac{\lambda_1}{2} = \frac{c}{2f_0}$. $N_m = 15$, $K = 3$, $f_0 = 100$ Hz, and the frequency set is $\{1, \dots, N_f\}\cdot f_0$ Hz. The $\ell_{1, 2}$ regularization parameter $\lambda = 0.6$ for (b). Each point represents 100 trials. The DOAs for each trial are randomly generated between $[10^\circ, 170^\circ]$ with with a minimum angular separation $4/N_m$. $\mathbf{x}_w \sim \mathcal{CN}(0, 1)$. }
    \label{rmse_plot_anm}
\end{figure}

We then change the spacing to $d = \frac{\lambda_{1}}{2}$ (See Fig. \ref{rmse_plot_anm}). In this case, aliasing and possible collisions will be present when $N_f \geq 2$. However, if more frequencies are available, such ambiguities can be potentially suppressed \cite{nannuru2019sparse}. For that reason, we only consider the case with $8$ frequencies from $100,\dots,800$ Hz. In Fig. \ref{rmse_plot_small} the frequencies were $12.5,\dots,100$ Hz, the aperture is here a factor 8 larger in Fig. \ref{rmse_plot_anm}. Although the error stops to decrease for ANM in the high SNR region due to the bias from the regularization, the performance still improves in the low SNR region if more frequencies are available. In addition, compared with Fig. \ref{rmse_plot_small} (d), the performance of ANM improves when SNR is between $0$ to $20$ dB, and that demonstrates the benefits of larger apertures.

\subsubsection{DOA Estimation under Varying \texorpdfstring{$K$}{Lg}}
We examine the DOA estimation performance under varying numbers of sources ($K$) in this section. Both the real flat (Fig.\ \ref{rmse_plot} (a)) and complex random amplitude source (Fig.\ \ref{rmse_plot} (b)) are tested under noise-free conditions. DOA is an integer randomly generated from a uniform distribution between [$0^\circ$, $180^\circ$]. Therefore, there is no grid mismatch issue for SBL. For the real and flat amplitude case ($\mathbf{x}_w = 1/\sqrt{N_f} \cdot \mathbf{1}_{N_f}$), ANM will be immune to collisions (or near collisions) since the fundamental constraint \eqref{dual_cons} and the dual certificate condition \eqref{cert} can be satisfied simultaneously. Therefore, the optimality is guaranteed and perfect DOA estimation is expected. In the complex random amplitude case,  since near collisions affect the performance of ANM, robust ANM (see \eqref{ssdp_robust}) is applied. 
From Fig. \ref{rmse_plot} (b), the DOA estimation error increases when the complex amplitude is applied for both methods. ANM (and robust ANM) still outperforms SBL for both real and complex amplitudes even if there is no grid mismatch for SBL. Fig. \ref{rmse_plot} (b) also demonstrates the effectiveness of robust ANM for suppressing near collisions. Because of the presence of near collisions in the complex amplitude case, more frequencies do not necessarily bring about better performance for ANM.

\subsubsection{DOA estimation under Varying DOA separation}
\begin{figure}
    \centering
    \includegraphics[width=8.5cm]{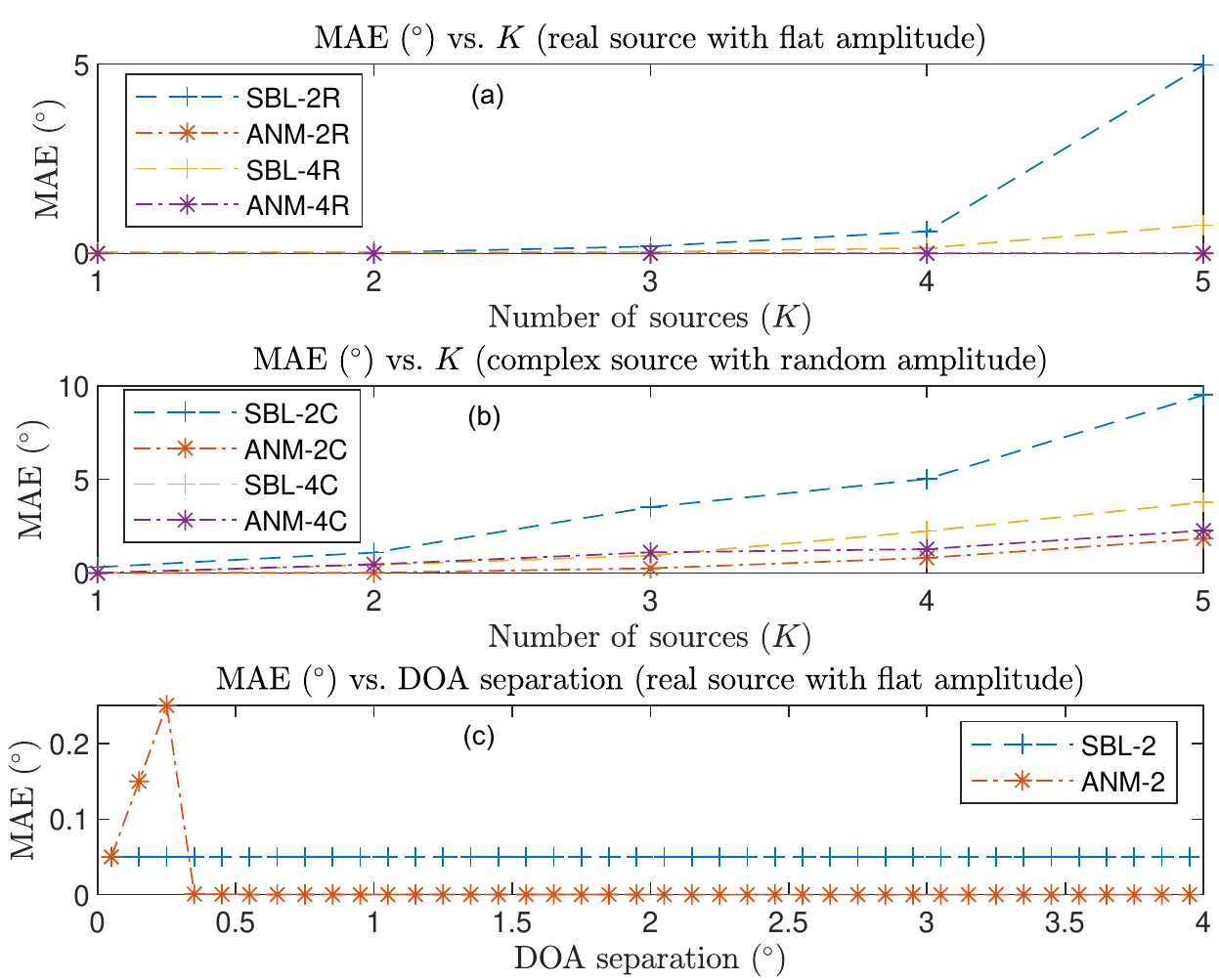}
    \caption{MAE ($^\circ$) vs. $K$ (a--b) and DOA separation (c). $N_m = 15$. Each point represents 100 trials, and no noise is present.  For (a--b), $N_f = \{2, 4\}$. For (b), robust ANM \eqref{ssdp_robust} is used. For (c), $N_f = 2$, $K = 2$, and $\mathbf{x}_w = 1/\sqrt{N_f} \cdot \mathbf{1}_{N_f}$. The first DOA is $90^\circ - $ DOA separation, and the second DOA is $90^\circ + $ DOA separation. The grid resolution for SBL is $0.1^\circ$. }
    \label{rmse_plot}
\end{figure}

Finally, we study the DOA estimation performance under different DOA separations. Since the amplitude is real and flat, ANM is immune to near collisions. From Fig. \ref{rmse_plot} (c), SBL has the same estimation error for all DOA separations and $N_f$. That error is entirely from the grid mismatch. However, the proposed gridless approach overcomes this issue and achieves exact DOA estimation.


\section{\label{sec:6}Conclusions}
The ANM framework is extended to support continuous parameter estimation across multiple frequencies. ANM is initially formulated as an equivalent SDP problem based on the bounded real lemma so that the ANM becomes computationally tractable. In addition, the dual certificate condition is derived. With the help of the dual certificate condition, the optimality can be certified, and the DOAs are identified by finding the roots of a polynomial. We also construct the dual certificate and show that a valid construction exists when the source amplitude has a uniform magnitude. Based on our signal model, the higher frequencies may have the risk of collision or near collision. These two cases are extensively studied and a robust ANM method with regularization is proposed for near collision suppression. The numerical results demonstrate the effectiveness of the proposed method.

\section*{Acknowledgement}
This work is supported by NSF Grant CCF-1704204, NSF Grant CCF-2203060, and Office of Naval Research (ONR) Grant N00014-21-1-2267. 

\bibliographystyle{IEEEtran}
\bibliography{refs, strings}

\appendix
\subsection{Proof for Proposition 3.1}
\label{proof_3.1}
Construct the Hermitian trigonometric polynomial
\begin{equation}
\label{R(w)}
    R(w) := 1 - \|\mathbf{H}^H\mathbf{z}\|_2^2  = 1 - \mathbf{z}^H\mathbf{H}\mathbf{H}^H\mathbf{z}.
\end{equation}
From \eqref{anmdual}, we know that $\|\mathbf{Q}\|^{*}_{\mathcal{A}}  \leq 1$ holds if and only if $R(w) \ge 0$ for all $w \in [-1/2, 1/2]$. 

First, suppose there exists a matrix $\mathbf{P}_0 \in \mathbb{C}^{N \times N} \succeq 0$ such that~\eqref{tr} and~\eqref{sdp} hold. We must argue that $R(w) \ge 0$ for all $w$. Consider the expression $\mathbf{z}^H\mathbf{P}_0\mathbf{z}$ and note that
\begin{align}
    \mathbf{z}^H\mathbf{P}_0\mathbf{z} &= \mathrm{Tr}(\mathbf{z}^H \mathbf{P}_0 \mathbf{z}) = \mathrm{Tr}(\mathbf{z}\mathbf{z}^H \mathbf{P}_0) = \sum_{k = -(N-1)}^{N-1}r_{k} z^{-k},\nonumber
\end{align}
 where $r_{k} = \sum_{i = 1}^{N-k}\mathbf{P}_0(i, i + k)$ for $k \geq 0$ and $r_{k} = r_{-k}^*$ for $k < 0$. 
From~\eqref{tr}, we  conclude that $\mathbf{z}^H\mathbf{P}_0\mathbf{z} = z^0 = 1$. Substituting this into $R(w)$ and defining  $\mathbf{P}_1 := \mathbf{H} \mathbf{H}^H$ gives
\[
R(w) = \mathbf{z}^H\mathbf{P}_0\mathbf{z}-\mathbf{z}^H\mathbf{P}_1\mathbf{z} = \mathbf{z}^H(\mathbf{P}_0-\mathbf{P}_1)\mathbf{z}.
\]
Since the matrix in~\eqref{sdp} is PSD, its Schur complement $\mathbf{P}_0  - \mathbf{H} \mathbf{I}_{N_f}^{-1} \mathbf{H}^H = \mathbf{P}_0 -\mathbf{P}_1 \succeq 0$, and so $R(w) \ge 0$ for all $w \in [-1/2, 1/2]$.

Next, suppose $R(w) \ge 0$ for all $w \in [-1/2, 1/2]$. We must argue that there exists a matrix $\mathbf{P}_0 \in \mathbb{C}^{N \times N} \succeq 0$ such that~\eqref{tr} and~\eqref{sdp} hold. Since $R(w) \ge 0$, $1 \ge \mathbf{z}^H\mathbf{P}_1\mathbf{z}$, where we have again defined $\mathbf{P}_1 := \mathbf{H} \mathbf{H}^H \succeq 0$. From~\cite[Lemma 4.25]{dumitrescu2017positive} and the fact that $1$ and $\mathbf{z}^H\mathbf{P}_1\mathbf{z}$ are univariate trigonometric polynomials, it follows that there exists $\mathbf{P}_0 \succeq \mathbf{P}_1$ such that $1 = \mathbf{z}^H\mathbf{P}_0\mathbf{z}$ and~\eqref{tr} hold.
The matrix in~\eqref{sdp} has Schur complement $\mathbf{P}_0  - \mathbf{H} \mathbf{I}_{N_f}^{-1} \mathbf{H}^H = \mathbf{P}_0 -\mathbf{P}_1 \succeq 0$, and therefore~\eqref{sdp} holds. $\hfill\square$

\subsection{Proof for Theorem 3.2}
\label{proof_3.2}
First, notice that if (\ref{cert}) is satisfied, based on (\ref{anmdual}), we have $\|\mathbf{Q}\|_{\mathcal{A}}^{*} \leq 1$. Then, 
\begin{align}
\|&\mathbf{X}\|_{\mathcal{A}} \geq \|\mathbf{X}\|_{\mathcal{A}} \cdot \|\mathbf{Q}\|_{\mathcal{A}}^{*}  
\overset{\text{(a)}}\geq \langle \mathbf{Q, X} \rangle_{\mathbb{R}} = \text{Re}[\mathrm{Tr}(\mathbf{Q}^H\mathbf{X})]\nonumber\\
&= \sum_{w \in \mathcal{W}}\text{Re}[\mathrm{Tr}(c_w\mathbf{Q}^H \mathbf{A}(w) \varoast \mathbf{x}^T_w)] \nonumber\\
&= \sum_{w \in \mathcal{W}}^{} \!\!\sum_{f = 1}^{N_f} \!\!\text{Re}[c_w  \mathbf{q}_{f}^H x_w(f)\mathbf{a}(f, w) ] 
=  \!\!\!\!\sum_{w \in \mathcal{W}}^{} \!\text{Re}[c_w \mathbf{x}_w^H\bm{\psi}(\mathbf{Q}, w) ] \nonumber\\
&\overset{\text{(b)}}= \sum_{w \in \mathcal{W}}^{} \text{Re}[c_w  \mathrm{sign}(c_w^*)\|\mathbf{x}_w\|_2^2 ]
= \sum_{w \in \mathcal{W}}^{} \!\!|c_w| \overset{\text{(c)}}\geq \|\mathbf{X}\|_{\mathcal{A}},
\end{align}
where (a) is based on H{\"o}lder's inequality, (b) follows because if $w \in \mathcal{W}$, then $\bm{\psi}(\mathbf{Q}, w) = \mathrm{sign}(c_w^*) \mathbf{x}_w$ based on (\ref{cert}), and (c) follows from the definition of the atomic norm \eqref{anm_def} as the infimum of the combination coefficients. 
Hence, $\|\mathbf{X}\|_{\mathcal{A}} = \langle \mathbf{Q, X} \rangle_{\mathbb{R}} = \sum_{w \in \mathcal{W}} |c_w|$.

For uniqueness, suppose there exists another decomposition $\mathbf{X} = \sum_{w'} c_{w'} \mathbf{A}(w') \varoast \mathbf{x}_{w'}^T$ which satisfies $\|\mathbf{X}\|_{\mathcal{A}} = \sum_{w'} |c_{w'}|$. There must exist $w' \notin \mathcal{W}$ contributing to $\mathbf{X}$ due to the mutual linear independence of the atoms. Therefore, we have the contradiction:
\begin{align}
&\sum_{w'} |c_{w'}| = \|\mathbf{X}\|_{\mathcal{A}} = \langle \mathbf{Q, X} \rangle_{\mathbb{R}} = \sum_{w'} \text{Re}[c_{w'} \langle \mathbf{x}_{w'}, \bm{\psi}(\mathbf{Q}, w') \rangle]\nonumber \\
&= \sum_{w' \in \mathcal{W}} \text{Re}[c_{w'} \mathbf{x}_{w'}^H\bm{\psi}(\mathbf{Q}, w')] +  \sum_{w' \notin \mathcal{W}} \text{Re}[c_{w'} \mathbf{x}_{w'}^H\bm{\psi}(\mathbf{Q}, w')] \nonumber\\
&\overset{\text{(a)}}< \sum_{w' \in \mathcal{W}} |c_{w'}| + \sum_{w' \notin \mathcal{W}} |c_{w'}| = \sum_{w'} |c_{w'}|,
\end{align}
where (a) is because of (\ref{cert}). Therefore, the atomic decomposition which satisfies $\|\mathbf{X}\|_{\mathcal{A}} = \sum_{w \in \mathcal{W}} |c_w|$ must be unique. $\hfill\square$

\subsection{Proof for Proposition \ref{p1_mbw}}
\label{proof_fast}
Construct the Hermitian trigonometric polynomial
\begin{equation}
\label{R(w)mbw}
    R(w) := 1 - \|\mathbf{H}_r^H\mathbf{z}_r\|_2^2  = 1 - \mathbf{z}_r^H\mathbf{H}_r\mathbf{H}_r^H\mathbf{z}_r.
\end{equation}
From \eqref{anmdual}, we know that $\|\mathbf{Q}\|^{*}_{\mathcal{A}}  \leq 1$ holds if and only if $R(w) \ge 0$ for all $w$. 

First, suppose there exists a matrix $\mathbf{P}_{r0} \in \mathbb{C}^{N_u \times N_u} \succeq 0$ such that~\eqref{tr_mbwr} and~\eqref{sdp_mbwr} hold. We must argue that $R(w) \ge 0$ for all $w$. Consider the expression $\mathbf{z}_r^H\mathbf{P}_{r0}\mathbf{z}_r$ and note that
\begin{align}
    \mathbf{z}_r^H\mathbf{P}_{r0}\mathbf{z}_r = \mathrm{Tr}(\mathbf{z}_r^H \mathbf{P}_{r0} \mathbf{z}_r) &= \mathrm{Tr}(\mathbf{z}_r\mathbf{z}_r^H \mathbf{P}_{r0}) = \!\!\!\!\!\!\!\!\sum_{k = -(N-1)}^{N-1}r_{k} z^{-k} \nonumber \\
    r_k &= \sum_{\substack{i, j, \mathcal{U}_j - \mathcal{U}_i = k}} \mathbf{P}_{r0}(i, j) \label{r_0mbwr}
\end{align}
for $k \geq 0$ and $r_{k} = r_{-k}^*$ for $k < 0$. From~\eqref{tr_mbwr}, we then conclude that $\mathbf{z}_r^H\mathbf{P}_{r0}\mathbf{z}_r = z^0 = 1$. Substituting this fact into $R(w)$ and defining  $\mathbf{P}_{r1} := \mathbf{H}_r \mathbf{H}_r^H$, we have
\begin{equation}
   R(w) = \mathbf{z}_r^H\mathbf{P}_{r0}\mathbf{z}_r-\mathbf{z}_r^H\mathbf{P}_{r1}\mathbf{z}_r = \mathbf{z}_r^H(\mathbf{P}_{r0}-\mathbf{P}_{r1})\mathbf{z}_r. 
\end{equation}

Since the matrix in~\eqref{sdp_mbwr} is PSD, its Schur complement $\mathbf{P}_{r0}  - \mathbf{H}_r \mathbf{I}_{N_f}^{-1} \mathbf{H}_r^H = \mathbf{P}_{r0} -\mathbf{P}_{r1} \succeq 0$, and so $R(w) \ge 0$ for all $w$. $\hfill\square$

\subsection{The Derivation of the Dual Problem of (20) } 
\label{dual_derive}

Consider the Lagrangian of \eqref{ssdp} given by
\begin{align}
     &\mathcal{L}(\mathbf{Q}, \mathbf{P}_0, \mathbf{H}, \mathbf{\Lambda}_1, \mathbf{\Lambda}_2, \mathbf{\Lambda}_3,  \mathbf{\Lambda}_Q, \mathbf{v}) = \nonumber\\
     & \langle \mathbf{Q}, \mathbf{Y} \rangle_{\mathbb{R}} - \bigg \langle \left[  
  \begin{array}{cc}   
    \mathbf{\Lambda}_1 & \mathbf{\Lambda}_2\\  
    \mathbf{\Lambda}_2^H & \mathbf{\Lambda}_3 \\  
  \end{array}
\right], \left[                 
  \begin{array}{cc}   
    \mathbf{P}_0 & \mathbf{H} \\  
    \mathbf{H}^H & \mathbf{I}_{N_f} \\  
  \end{array}
\right] \bigg \rangle_{\mathbb{R}} \nonumber\\
     &- \sum_{k = 0}^{N-1}v_k (\delta_k - \sum_{j - i = k}\mathbf{P}_0(i, j)) - \langle \mathbf{\Lambda}_Q, \mathbf{H} - \mathcal{R}^*(\mathbf{Q}) \rangle_{\mathbb{R}} \nonumber\\
     &= \mkern-5mu \langle \mathbf{Q}, \mathbf{Y} \rangle_{\mathbb{R}} \mkern-5mu + \mkern-5mu \langle \mathbf{\Lambda}_Q, \mathcal{R}^*(\mathbf{Q}) \rangle_{\mathbb{R}} \mkern-5mu - \mkern-5mu [\langle \mathbf{P}_0, \mathbf{\Lambda}_1  \rangle_{\mathbb{R}} \mkern-5mu + \mkern-5mu 2 \langle \mathbf{\Lambda}_2, \mathbf{H} \rangle_{\mathbb{R}} \mkern-5mu + \mkern-5mu \mathrm{Tr}(\mathbf{\Lambda}_3)] \nonumber\\
     &- \mathbf{v}_0 + \langle \mathbf{P}_0 , \mathrm{Toep}(\mathbf{v}) \rangle_{\mathbb{R}} - \langle \mathbf{\Lambda}_Q, \mathbf{H} \rangle_{\mathbb{R}}.
\end{align}
The derivation uses: $ \sum_{k = 0}^{N-1}v_k \sum_{j - i = k}\mathbf{P}_0(i, j) = \langle \mathbf{P}_0 , \mathrm{Toep}(\mathbf{v}) \rangle_{\mathbb{R}}$. Further, the dual matrix $\left[                 
  \begin{array}{cc}   
    \mathbf{\Lambda}_1 & \mathbf{\Lambda}_2\\  
    \mathbf{\Lambda}_2^H & \mathbf{\Lambda}_3 \\  
  \end{array}
\right]$ associated with the inequality constraint $\left[                 
  \begin{array}{cc}   
    \mathbf{P}_0 & \mathbf{H} \\  
    \mathbf{H}^H & \mathbf{I}_{N_f} \\  
  \end{array}
\right]  \succeq 0$ is an PSD matrix to ensure  the inner product between these two matrices is non-negative, whereby  the optimal value for the dual problem gives a lower bound for the primal problem.

The dual function is 
%
\begin{align}
 g(\mathbf{\Lambda}_1, \mkern-3mu \mathbf{\Lambda}_2, \mkern-3mu \mathbf{\Lambda}_3, \mkern-3mu \mathbf{\Lambda}_Q, \mkern-3mu \mathbf{v}) &\mkern-6mu = \mkern-6mu \inf_{\mathbf{Q}, \mkern-3mu\mathbf{P}_0, \mkern-3mu \mathbf{H}} \mathcal{L}(\mathbf{Q}, \mathbf{P}_0, \mathbf{H}, \mathbf{\Lambda}_1, \mathbf{\Lambda}_2, \mathbf{\Lambda}_3, \mathbf{\Lambda}_Q, \mathbf{v}) \nonumber\\
 & \mathrm{s.t.} \left[                 
  \begin{array}{cc}   
    \mathbf{\Lambda}_1 & \mathbf{\Lambda}_2\\  
    \mathbf{\Lambda}_2^H & \mathbf{\Lambda}_3 \\  
  \end{array}
\right] \succeq 0.\label{dual_fun}
\end{align}

The infimum of $\mathcal{L}$ over $\mathbf{Q}$ is thereby $\mtight \inf_{\mathbf{Q}} J(\mathbf{Q}) := [\langle \mathbf{Q}, \mathbf{Y}  \rangle_{\mathbb{R}} + \langle \mathbf{\Lambda}_Q, \mathcal{R}^*(\mathbf{Q}) \rangle_{\mathbb{R}}] = [\langle \mathbf{Y}, \mathbf{Q}  \rangle_{\mathbb{R}} + \langle \mathcal{R}(\mathbf{\Lambda}_Q), \mathbf{Q} \rangle_{\mathbb{R}}] = \langle \mathbf{Y} + \mathcal{R}(\mathbf{\Lambda}_Q), \mathbf{Q}  \rangle_{\mathbb{R}}$. The infimum of $J(\mathbf{Q})$ is bounded only if $\mathbf{Y} = -\mathcal{R}(\mathbf{\Lambda}_Q)$. Similarly, the infimum of $\mathcal{L}$ over $\mathbf{P}_0$ is bounded only if $\mathrm{Toep}(\mathbf{v}) =  \mathbf{\Lambda}_1 \succeq 0.$ The infimum of $\mathcal{L}$ over $\mathbf{H}$ is bounded only if $\mathbf{\Lambda}_Q = -2\mathbf{\Lambda}_2$. Consider $2\mathbf{\Lambda}_2 = \widetilde{\mathbf{Y}}$, then we must have $\mathbf{Y} = -\mathcal{R}(\mathbf{\Lambda}_Q) = \mathcal{R}(2\mathbf{\Lambda}_2) = \mathcal{R}(\widetilde{\mathbf{Y}})$.

Consider $\mathbf{\Lambda}_3 = \frac{1}{2}\mathbf{W}$, and $\mathbf{v} = \frac{1}{2}\mathbf{u}$,  the dual function becomes  $-\frac{1}{2}\mathrm{Tr}(\mathbf{W}) -\frac{1}{2}\mathrm{Tr}(\mathrm{Toep}(\mathbf{u})).$ The dual problem is
\begin{align}
     \max_{\mathbf{W}, \mathbf{u}, \mathbf{\widetilde{Y}}} &\mkern-5mu -\frac{1}{2}[\mathrm{Tr}(\mathbf{W}) \mkern-5mu + \mkern-5mu \mathrm{Tr(Toep}(\mathbf{u}))] 
    \mkern-5mu \quad \mkern-5mu 
    \nonumber \\
  &  \textrm{s.t.} \mkern-4mu
\left[\mkern-7mu                 
  \begin{array}{cc}   
    \mathrm{Toep}(\mathbf{u}) \mkern-7mu & \mkern-7mu \mathbf{\widetilde{Y}} \\  
    \mathbf{\widetilde{Y}}^H \mkern-7mu & \mkern-7mu \mathbf{W} \\  
  \end{array}
\mkern-7mu \right] \mkern-5mu \succeq 0, \mkern-4mu \mathbf{Y} \mkern-6mu = \mkern-6mu \mathcal{R}(\mathbf{\widetilde{Y}}),
\end{align}
which is equivalent to~\eqref{ssdp_dual_2}. $\hfill\square$

\subsection{Properties for Exact Collision}
\label{appendix_collision}
\subsubsection{\texorpdfstring{$\mathbf{K}_i$}{Lg} is Singular}
\label{singular_K}
First observe that $\mathbf{K}_i$ in \eqref{Ki} is singular. We also  recognize the periodicity of $K_i(w)$. Since $K_i(w) = K_i(w+k/i) (k < i, i \in \{1, \dots, N_f \})$, $k/i$ is the period for $K_i(w)$. In addition, $k/i$ is also the period for ${K}'_i(w)$ and ${K}''_i(w)$. Without loss of generality, we assume there exists collision between $w_1$ and $w_2$ (i.e. ${|w_1 - w_2| = \frac{k}{i}}$), then \begin{equation}
\label{K_period}
\begin{aligned}
&{K}_i(0) = {K}_i(w_1 - w_2) = {K}_i(w_2 - w_1) = 1,      \\
& {K}'_i(0) = {K}'_i(w_1 - w_2) = {K}'_i(w_2 - w_1) = 0, ~\textrm{and}     \\
& {K}''_i(0) = {K}''_i(w_1 - w_2) = {K}''_i(w_2 - w_1).
\end{aligned}
\end{equation}
The first and second row of $\mathbf{K}_i$ are
\begin{align}
    &[K_i(w_1 \!\! - \!\!w_1) \dots K_i(w_1 \!\! - \!\! w_K) \dots K_i'(w_1 \!\! - \!\! w_1) \dots K_i'(w_1 \!\! - \!\! w_K)] \nonumber \\
    &[K_i(w_2 \!\!- \!\! w_1) \dots K_i(w_2 \!\! - \!\! w_K) \dots K_i'(w_2 \!\! - \!\! w_1) \dots K_i'(w_2 \!\! - \!\! w_K)].
\end{align}
Note that $K_i(w_2 - w_j) = K_i(w_1 - w_j - (w_1 - w_2)) = K_i(w_1 - w_j)$ and $K_i'(w_2 - w_j) = K_i'(w_1 - w_j - (w_1 - w_2)) = K_i'(w_1 - w_j)$ for any $j$.
%
Thus, the first two rows are identical. $\mathbf{K}_i$ is hence rank-deficient and singular.  

However, the singularity of $\mathbf{K}_i$ does not imply that the solution to the system of equations (\ref{linear}) does not exist. If $\left[\begin{tabular}{@{}c@{}}
$\mathrm{sign}(c_w^{*})x_{w_1}(i)$ \dots $\mathrm{sign}(c_w^{*})x_{w_{K}}(i)$  0 \dots 0
\end{tabular}
\right]^T := \widehat{\mathbf{x}_i}$ lies in the range space of $\mathbf{K}_i$, the solution of \eqref{linear} exists but non-unique. Among the infinite number of solutions, we choose the Moore-Penrose pseudoinverse solution $\mathbf{K}_i^\dagger \widehat{\mathbf{x}_i}$. 

\subsubsection{Recovery for the Coefficients not Possible}\label{recovery}
Here, we  discuss the possibility of recovering the coefficients under the collision condition. Although it is possible to localize the sources, the recovery of the coefficients $\widehat{c}_k \widehat{\mathbf{x}}_k$ is not possible due to the fundamental limit in (\ref{aliasing_manifold}).

The DOAs are localized by finding the peak of the dual polynomial vector under the collision condition. For the estimated DOAs $(\widehat{w}_1, ..., \widehat{w}_{K})$, (\ref{KR}) gives
\begin{equation}
    \mathbf{X} = \sum_{k = 1}^{K} \widehat{c}_k \mathbf{A}(\widehat{w}_k) \varoast \widehat{\mathbf{x}}_k^T = \sum_{k = 1}^{K} \mathbf{A}(\widehat{w}_k) \varoast \widetilde{\mathbf{x}}_k^T,
\end{equation}
where $\mtight \widetilde{\mathbf{x}}_k := \widehat{c}_k \widehat{\mathbf{x}}_k$. Since $\mtight \mathbf{Y} = \mathbf{X} = [\mathbf{y}_1  ... \mathbf{y}_{N_f}]$, the entries in $\widetilde{\mathbf{x}}_k$ are recovered by solving $\mtight \mathbf{y}_{f} = \sum_{k = 1}^{K}\mathbf{a}(f, \widehat{w}_k)\widetilde{\mathbf{x}}_k = [\mathbf{a}(f, \widehat{w}_1)  ...  \mathbf{a}(f, \widehat{w}_{K})][\widetilde{\mathbf{x}}_1(f)  ...  \widetilde{\mathbf{x}}_{K}(f)]^T (f  = 1,\dots, N_f)$.

However, when $f = i$,  $\mathbf{a}(i, \widehat{w}_1) = \mathbf{a}(i, \widehat{w}_2)$ from (\ref{aliasing_manifold}). Then, $\mathbf{a}(i, \widehat{w}_1)\widetilde{\mathbf{x}}_1(i) +  \mathbf{a}(i, \widehat{w}_2)\widetilde{\mathbf{x}}_2(i) = \mathbf{a}(i, \widehat{w}_1)[\widetilde{\mathbf{x}}_1(i) + \widetilde{\mathbf{x}}_2(i)]$. Therefore, we have to decouple $\widetilde{\mathbf{x}}_1(i)$ and $\widetilde{\mathbf{x}}_2(i)$ based on their sum, which is impossible.


\subsection{Proof for Lemma \ref{bound_alpha_beta}}
\label{bound_lemma}
 %
 From below \eqref{cert_new}, $ |\overline{\mathbf{x}}_{w}(i)| =|\mathbf{x}_{w_{2}}(i)|$,
 we have
 \begin{align}
 \|{\alpha}_i\|_{\infty} &= \bigg\| \mathbf{S}_i^{-1} \left[\begin{matrix}
 \mathrm{sign}(c_w^{*})\overline{\mathbf{x}}_{w_1}(i) \\ \mathrm{sign}(c_w^{*})\overline{\mathbf{x}}_{w_{2}}(i)\end{matrix}
 \right] \bigg\|_{\infty} 
  \nonumber \\
  &\leq \| \mathbf{S}_i^{-1} \|_{\infty} \bigg\| \left[\begin{matrix}
  \mathrm{sign}(c_w^{*})\mathbf{x}_{w_1}(i) \\ \mathrm{sign}(c_w^{*})\mathbf{x}_{w_{2}}(i)\end{matrix}
 \right] \bigg\|_{\infty} \!\!\! \leq \|\mathbf{S}_i^{-1} \|_{\infty},   \label{alpha_bound1}
\\
 \|{\beta}_i\|_{\infty}\!\! &\leq \bigg\|  \mathbf{D}_{i, 2}^{-1}\mathbf{D}_{i, 1}\mathbf{S}_i^{-1} \left[\begin{matrix}
 \mathrm{sign}(c_w^{*})\mathbf{x}_{w_1}(i) \\ \mathrm{sign}(c_w^{*})\mathbf{x}_{w_{2}}(i)\end{matrix}
 \right] \bigg\|_{\infty} \nonumber \\
& \!\! \!\! \!\! \!\! \leq \|\mathbf{D}_{i, 2}^{-1}\mathbf{D}_{i, 1}\mathbf{S}_i^{-1} \|_{\infty} \leq \! \| \mathbf{D}_{i, 2}^{-1}\|_{\infty}\| \!\mathbf{D}_{i, 1}\|_{\infty} \!\|\mathbf{S}_{i}^{-1}\|_{\infty}.  \label{beta_bound1}
\end{align}
%
$\|\mathbf{S}_i^{-1}\|_{\infty}$ is bounded as
\begin{align}
     \|\mathbf{S}_i^{-1}&\|_{\infty} = \|(\mathbf{D}_{i, 0} - \mathbf{D}_{i, 1}\mathbf{D}_{i, 2}^{-1}\mathbf{D}_{i, 1})^{-1}\|_{\infty} \nonumber\\
     &\leq {1}{/(1 - \|\mathbf{I} - (\mathbf{D}_{i, 0} - \mathbf{D}_{i, 1}\mathbf{D}_{i, 2}^{-1}\mathbf{D}_{i, 1}) \|_{\infty})} \nonumber\\
     &\leq {1}{/[1 - (\|\mathbf{I} - \mathbf{D}_{i, 0}\|_{\infty} + \|\mathbf{D}_{i, 1} \|_{\infty}^2 \|\mathbf{D}_{i, 2}^{-1}\|_{\infty})]}.
\end{align}


Inspired by the proof of   \cite[Lemma 2.2]{candes2014towards}, the  bounds for $\| \mathbf{I} - \mathbf{D}_{i, 0}\|_{\infty}$, $\|\mathbf{D}_{i, 1}\|_{\infty}$, and $\| K_i''(0)\mathbf{I} - \mathbf{D}_{i, 2} \|_{\infty}$ are established (define $d_0 := 6.253 \times 10^{-3}, d_1 := 7.639 \times 10^{-2}, d_2 := 1.053, d_3 := 11/32 \pi^2 $, where $d_0, d_1,$ and $d_2$ are empirical \cite{candes2014towards} and $d_3$ is analytical):
\begin{align}
& \| \mathbf{I} \mkern-4mu - \mkern-4mu \mathbf{D}_{i, 0}\|_{\infty} \leq \Big \|\mathbf{I} \mkern-4mu - \mkern-4mu \frac{1}{i} \mathbf{I} \Big \|_{\infty} \mkern-6mu + \mkern-6mu \Big \| \frac{1}{i} \mathbf{I} \mkern-4mu - \mkern-4mu \mathbf{D}_{i, 0}  \Big \|_{\infty} \mkern-7mu = \mkern-5mu 1 \mkern-5mu - \mkern-5mu \frac{1}{i} \mkern-5mu + \mkern-5mu |K_i (w_1 \mkern-7mu - \mkern-7mu w_2)| \nonumber \\
    & = 1 \mkern-5mu - \mkern-5mu \frac{1}{i} + \frac{1}{i}|K_1(i (w_1 - w_2) )| \leq 1 \mkern-5mu + \mkern-5mu \frac{d_0 - 1}{i},\nonumber 
\\
&\| \mathbf{D}_{i, 1} \|_{\infty} = |K_i'(w_1 - w_2)| = |K_1'(i(w_1 - w_2))| \leq d_1 f_c,\nonumber 
\\
&\|K_i''(0)\mathbf{I} \mkern-6mu - \mkern-6mu \mathbf{D}_{i, 2} \|_{\infty} \mkern-6mu = \mkern-6mu |K_i''(w_1 \mkern-6mu - \mkern-6mu w_2)| \mkern-6mu = \mkern-6mu i |K_1''(i(w_1 \mkern-6mu - \mkern-6mu w_2))| \leq i  d_2 f_c^2,\label{D_2}\nonumber
\\
&|{K}_i''(0)| = \frac{i \pi^2f_c(f_c +4)}{3} \geq \frac{i\pi^2f_c^2}{3} + \frac{4i \pi^2 f_c^2}{3 \cdot 128} = i \cdot d_3  f_c^2. 
\end{align}
Therefore, $\|\mathbf{D}_{i, 2}^{-1}\|_{\infty}$ is bounded as follows ($d_4 := 1/(d_3 - d_2) = 0.4275$)
\begin{equation}
\label{bound_i2}
\begin{aligned}
     \| \mathbf{D}_{i, 2}^{-1} \|_{\infty} \leq \frac{1}{|K_i''(0)| \mkern-6mu - \mkern-6mu \| K_i''(0)\mathbf{I} \mkern-6mu - \mkern-6mu \mathbf{D}_{i, 2} \|_{\infty}} \mkern-6mu \leq \mkern-6mu \frac{1}{i (d_3 \mkern-6mu - \mkern-6mu d_2) f_c^2} \mkern-6mu = \mkern-6mu \frac{d_4}{i f_c^2}.
\end{aligned}
\end{equation}

Then, following \eqref{alpha_bound1} and \eqref{beta_bound1}, the bounds for $\|{\alpha}_i\|_{\infty}$ and $\|{\beta}_i\|_{\infty}$ are (define $c_\alpha := 1.008824$, and $c_\beta := 3.294 \times 10^{-2}$):
\begin{align}
    \|{\alpha}_i\|_{\infty} &\leq \|\mathbf{S}_i^{-1} \|_{\infty} \leq \frac{i}{1 - d_0 - d_1^2 d_4} := i \cdot c_\alpha ,
\\
\|{\beta}_i\|_{\infty}  &\leq \| \mathbf{D}_{i, 2}^{-1}\|_{\infty}\| \mathbf{D}_{i, 1}\|_{\infty} \|\mathbf{S}_{i}^{-1}\|_{\infty} \mkern-5mu \leq \mkern-5mu \frac{d_1d_4}{f_c(1 \mkern-5mu - \mkern-5mu d_0 \mkern-5mu- \mkern-5mu d_1^2 d_4)} \nonumber \mkern-5mu := \mkern-5mu\frac{c_\beta}{f_c}.
\end{align}
$\hfill\square$

\subsection{Invertibility of $\mathbf{K}_i$}\label{invert_K}
\textcolor{black}{
Using the Schur complement, $\mathbf{K}_i$ is invertible 
if $\mathbf{D}_{i, 2}$ and the Schur complement $\mathbf{S}_i := \mathbf{D}_{i, 0} - \mathbf{D}_{i, 1}\mathbf{D}_{i, 2}^{-1}\mathbf{D}_{i, 1}$ are both invertible. To show that, we use the fact that a Hermitian matrix $\mathbf{M}$ is invertible if $\|\mathbf{I} - \mathbf{M} \|_{\infty} < 1$ \cite[eq. (2.12)]{candes2014towards}. }

\textcolor{black}{We begin with $\mathbf{D}_{i, 2}$. 
Notice $|K_i''(0)| = i|K_1''(0)| =   \frac{i \pi^2f_c(f_c +4)}{3} $. Therefore, based on \eqref{D_2},
\begin{equation}
\label{D_i2}
    \Big\|\mathbf{I}  -  \frac{\mathbf{D}_{i, 2}}{K_i''(0)} \Big\|_{\infty} \!\!\!\!\! = \! \frac{\|K_i''(0)\mathbf{I} \! - \! \mathbf{D}_{i, 2} \|_{\infty}}{|K_i''(0)|} \leq \frac{i  d_2 f_c^2}{i  \pi^2 f_c(f_c + 4) / 3} \! < \! 1,
\end{equation}
which implies that $\frac{\mathbf{D}_{i, 2}}{K_i''(0)}$ is invertible. Hence, $\mathbf{D}_{i, 2}$ is also invertible. }
\textcolor{black}{ We then consider the invertibility of $\mathbf{S}_i$. Based on the triangle inequality, 
\begin{equation}
\label{trian_eq}
    \|\mathbf{I} - \mathbf{S}_i \|_{\infty}  \leq \|\mathbf{I} - \mathbf{D}_{i, 0} \|_{\infty} + \|\mathbf{D}_{i, 1} \|_{\infty}^2 \|\mathbf{D}_{i, 2}^{-1}\|_{\infty}.
\end{equation}
Hence, to show $\|\mathbf{I} - \mathbf{S}_i \|_{\infty} < 1$, $\|\mathbf{I} - \mathbf{D}_{i, 0}\|_{\infty}$, $\| \mathbf{D}_{i, 1} \|_{\infty}$, and $\| \mathbf{D}_{i, 2}^{-1} \|_{\infty}$ need to be bounded.
}

\textcolor{black}{Plugging in the bounds in \eqref{D_2}, and \eqref{bound_i2}, we have 
\begin{equation}
\begin{aligned}
\label{S_i}
    \|\mathbf{I} \!\! - \!\! \mathbf{S}_i \|_{\infty} \leq 1 \! + \! \frac{d_0 \!\! + \!\! d_1^2 d_4 - 1}{i} = 1 \!\! + \!\! \frac{8.747 \times 10^{-3} \! - \! 1}{i} < 1,
\end{aligned}
\end{equation}
which implies that $\mathbf{S}_i$ is invertible. $\hfill\square$}

\subsection{Proof for Theorem \ref{case2_theorem}}
\label{appendix_proof}
For simplicity, we assume $K = 2$ in this section. But the theorem can be generalized to $K \geq 2$ if the separation condition is satisfied. Based on the assumption  $|\mathbf{x}_{w_1}(i)| = |\mathbf{x}_{w_2}(i)| = 1/\sqrt{N_f}$ for $\forall i \in \{1, ..., N_f \}$, as long as each entry in the constructed dual polynomial vector satisfies $|\bm{\psi}^i(w; w_1, w_2)| < 1/\sqrt{N_f}$, then $\|\bm{\psi}(w)\|_2 < 1$. Therefore, the bounds in Lemma 4.3 (2) further indicate $|\bm{\psi}^i(w; w_1, w_2)|$ (denote  $c_{\alpha}:= 1.008824$,  $c_{\beta} := 3.294 \times 10^{-2}$, $c := \frac{1}{\sqrt{N_f}}$)
\begin{equation}
\label{psi_i}
\begin{aligned}
    &|\bm{\psi}^i(w; w_1, w_2)| = |\!\!\! \sum_{k \in \{1, 2\}} \!\!\!{\alpha}_{k, i}{K}_i(w - w_k) + \! {\beta}_{k, i}{K}'_i(w - w_k)| \\
    &\leq \|{\alpha}_i \|_{\infty} \!\!\! \sum_{k \in \{1, 2\}} \!\!\! |{K}_i(w - w_k)| + \|{\beta}_i \|_{\infty} \sum_{k \in \{1, 2\}} |{K}_i'(w - w_k)| \\
    &\leq c[i c_{\alpha} \!\!\! \sum_{k \in \{1, 2\}} \!\!\! \frac{|K_1(i(w - w_k))|}{i} + \frac{c_{\beta}}{f_c} \!\! \sum_{k \in \{1, 2\}} \!\! |K_1'(i(w - w_k))|] \\
    &= c[c_{\alpha} \!\! \sum_{k \in \{1, 2\}} \!\! |K_1(i(w - w_k))| + \frac{c_{\beta}}{f_c} \! \sum_{k \in \{1, 2\}} \!\! |K_1'(i(w - w_k))|] \\
    &=\! c[\!\!\!\!\sum_{k \in \{1, 2\}}\!\!\!\!c_{\alpha} |K_1(i(w \!- \!w_k) \!\!\!\!\mod 1)| \! + \! \frac{c_{\beta}}{f_c} |K_1'(i(w \!- \! w_k) \!\!\!\!\mod 1)|].
\end{aligned}
\end{equation}

When $i = 1$,   
\begin{equation}
\label{psi_1}
\begin{aligned}
     |\bm{\psi}^1(w; w_1,\! w_2)| \! \leq \! c[c_{\alpha} \!\!\!\! \sum_{k \in \{1, 2\}} \!\!\! |K_1(w \mkern-6mu - \mkern-6mu w_k)| \mkern-4mu + \mkern-4mu \frac{c_{\beta}}{f_c} \!\! \sum_{k \in \{1, 2\}}\!\!\! |K_1'(w \mkern-6mu - \mkern-6mu w_k)|].
\end{aligned}
\end{equation}
We show $c[c_{\alpha} \sum_{k \in \{1, 2\}} |K_1(w - w_k)| + \frac{c_{\beta}}{f_c}  \sum_{k \in \{1, 2\}} |K_1'(w - w_k)|] < \frac{1}{\sqrt{N_f}}$ by applying \cite[Lemma 2.3 and 2.4]{candes2014towards}. We consider both the near and far regions. The near region $\mathcal{T}_{\text{near}}$ and far region $\mathcal{T}_{\text{far}}$ are defined as $\mathcal{T}_{\text{near}}:= \cup_{k = 1}^2[w_k - \nu, w_k + \nu]$ and $\mathcal{T}_{\text{far}}:= [0, 1] \backslash \mathcal{T}_{\text{near}}$, where $\nu = \frac{0.1649}{f_c}$. 

For $\mathcal{T}_{\text{far}}$, based on  \cite[Lemma 2.4]{candes2014towards}
\begin{equation}
\begin{aligned}
     &c_{\alpha} \sum_{k \in \{1, 2\}} |K_1(w - w_k)| + \frac{c_{\beta}}{f_c} \sum_{k \in \{1, 2\}} |K_1'(w - w_k)| \\
     &\leq 0.99992 < 1. 
\end{aligned}
\end{equation}
Therefore, 
\begin{equation}
\begin{aligned}
      |\bm{\psi}^1(w; w_1, w_2)| \leq c[c_{\alpha} \mkern-11mu \sum_{k \in \{1, 2\}} \mkern-10mu |K_1(w \mkern-6mu - \mkern-6mu w_k)| \mkern-6mu + \mkern-6mu \frac{c_{\beta}}{f_c} \mkern-9mu \sum_{k \in \{1, 2\}} \mkern-11mu |K_1'(w \mkern-6mu- \mkern-6mu w_k)|] \mkern-6mu < \mkern-6mu c . \nonumber
\end{aligned}
\end{equation}
If $i > 1$, the only difference between the last line of (\ref{psi_i}) and the right hand side of (\ref{psi_1}) is the dilation of $K_1$ and $K_1'$. This indicates the $i$-th entry is a special case for $i = 1$. Therefore, $|\bm{\psi}^i(w; w_1, w_2)| < c = 1/{\sqrt{N_f}}$ will also hold for $i > 1$. Hence, in $\mathcal{T}_{\text{far}}$, $\|\bm{\psi}(w)\|_2 < 1$ for $w \notin \mathcal{W}$.

For $\mathcal{T}_{\text{near}}$, inspired by the proof  in \cite[Lemma 2.3]{candes2014towards}, we show the strict concavity of $|\bm{\psi}^i(w; w_1, w_2)|$. We have
\begin{equation}
\begin{aligned}
     &\bm{\psi}^i_{R}(w)\bm{\psi}^{i''}_{R}(w) + |\bm{\psi}^{i'}(w)|^2 + |\bm{\psi}^i_{I}(w)||\bm{\psi}^{i''}_{I}(w)| \\
     &\leq -9.291 \times 10^{-2}(if_c / \sqrt{N_f})^2 < 0
\end{aligned}
\end{equation}
and
\begin{equation}
\begin{aligned}
     &\frac{\mathrm{d}^2|\bm{\psi}^i|(w)}{\mathrm{d} w^2} = -\frac{(\bm{\psi}^i_{R}(w)\bm{\psi}^{i'}_{R}(w) + \bm{\psi}^i_{I}(w)\bm{\psi}^{i'}_{I}(w))^2}{|\bm{\psi}^i(w)|^3} \\
     &+ \frac{\bm{\psi}^i_{R}(w)\bm{\psi}^{i''}_{R}(w) + |\bm{\psi}^{i'}(w)|^2 + |\bm{\psi}^i_{I}(w)||\bm{\psi}^{i''}_{I}(w)| }{|\bm{\psi}^i(w)|} < 0.
\end{aligned}
\end{equation}

Since $\bm{\psi}^{i'}(w_1) = \bm{\psi}^{i'}(w_2) = 0$, local strict concavity will imply $|\bm{\psi}^i(w;w_1, w_2)| < 1/\sqrt{N_f}$ in $\mathcal{T}_{\text{near}}$. $\hfill\square$

\begin{IEEEbiography}
[{\includegraphics[width=1in,height=1.25in,clip,keepaspectratio]{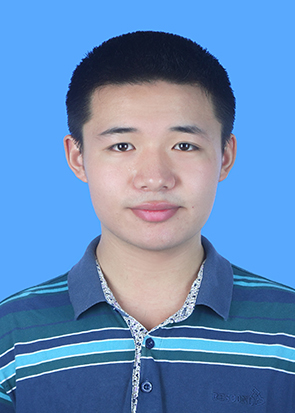}}]{Yifan Wu} received the M.S. from the University of California, San Diego in 2021. He is now pursuing Ph.D. degree in the University of California, San Diego. His research interest includes statistical signal processing, machine learning, and optimization theory. 
\end{IEEEbiography}

\begin{IEEEbiography}
[{\includegraphics[width=1in,height=1.25in,clip,keepaspectratio]{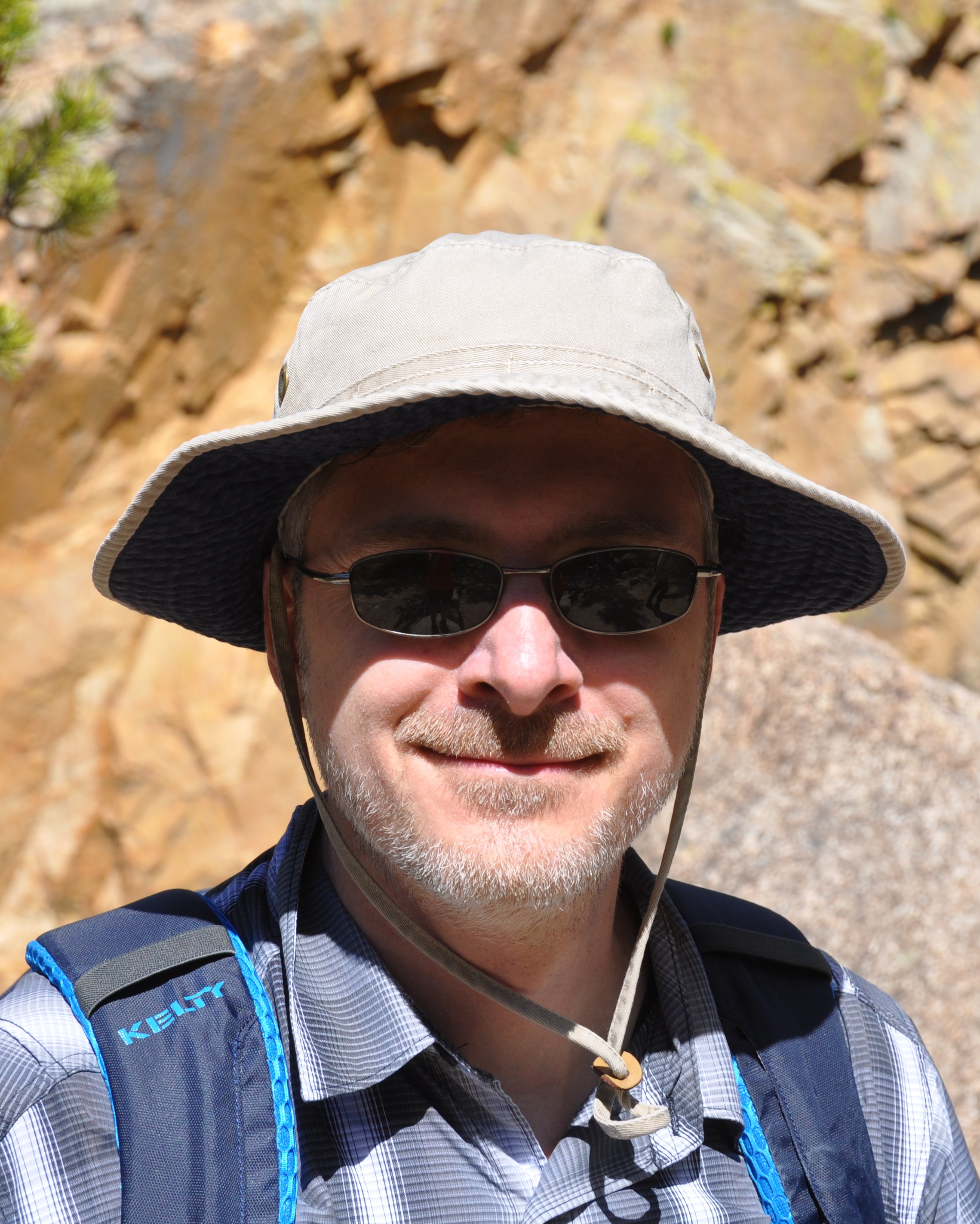}}]{Michael B. Wakin} (Fellow, IEEE) is a Professor of Electrical Engineering at the Colorado School of Mines. Dr. Wakin received a Ph.D. in electrical engineering in 2007 from Rice University. He was an NSF Mathematical Sciences Postdoctoral Research Fellow at Caltech from 2006-2007, an Assistant Professor at the University of Michigan from 2007-2008, and a Ben L. Fryrear Associate Professor at Mines from 2015-2017. His research interests include signal and data processing using sparse, low-rank, and manifold-based models.

In 2008, Dr. Wakin received the DARPA Young Faculty Award for his research in compressive multi-signal processing for environments such as sensor and camera networks. In 2012, Dr. Wakin received the NSF CAREER Award for research into dimensionality reduction techniques for structured data sets. Dr. Wakin is a recipient of the Best Paper Award and the Signal Processing Magazine Best Paper Award from the IEEE Signal Processing Society. He has served as an Associate Editor for IEEE Signal Processing Letters and IEEE Transactions on Signal Processing, and he is currently a Senior Area Editor for IEEE Transactions on Signal Processing.
\end{IEEEbiography}

\begin{IEEEbiography}
[{\includegraphics[width=1in,height=1.25in,clip,keepaspectratio]{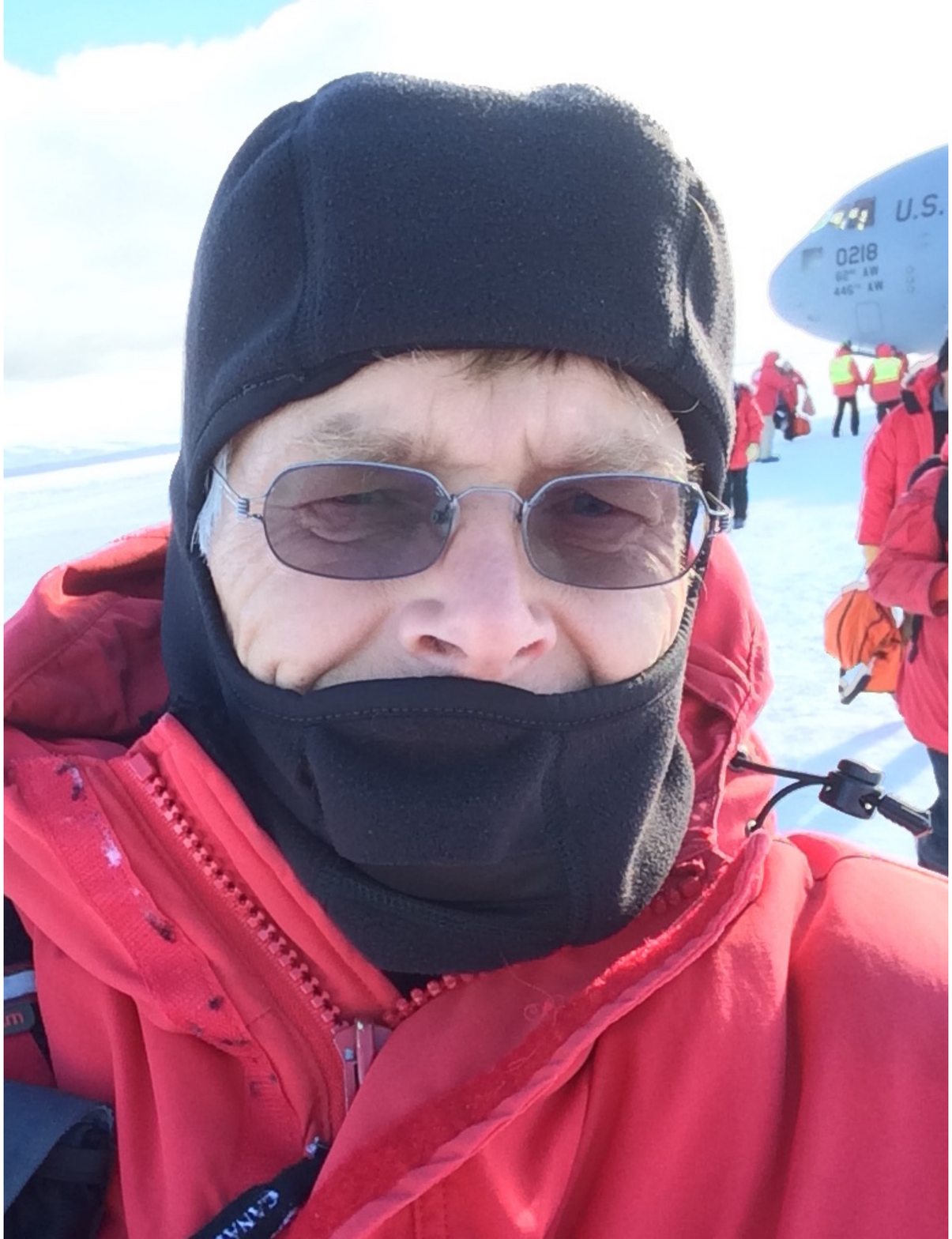}}]{Peter Gerstoft} (Fellow, IEEE) received the Ph.D. from  the Technical University  of Denmark, Lyngby, Denmark, in 1986. Since 1997, he has been with the University of California, San Diego. His current research interests are signal processing and machine leaning applied to acoustic, seismic, and electromagnetic signals. For more information see \url{http://noiselab.ucsd.edu}.
\end{IEEEbiography}

\ifCLASSOPTIONcaptionsoff
  \newpage
\fi

\newpage

\end{document}